\pdfoutput=1 
\documentclass[ 
 aps,%
 twocolumn,%
 reprint,%
 prl,%
 groupedaddress,%
 superscriptaddress,%
 showpacs,%
 lengthcheck,%
 letterpaper,%
 amsfonts,%
 floatfix,%
 citeautoscript
]{revtex4-1} 

\RequirePackage{amsmath,amssymb,bm}
\RequirePackage{graphicx}
\RequirePackage[normalem]{ulem}
\RequirePackage[]{subfigure}
\RequirePackage{hyperref}
\hypersetup{%
  breaklinks = {true},
  citecolor = {blue},
  colorlinks = {true},
  linkcolor = {red},
  pdfauthor = {\textcopyright\ F\'{a}bio Hip\'{o}lito },
  pdfcreator = {\LaTeX\ and \flqq hyperref\frqq},
  pdffitwindow = {true},
  pdfmenubar = {true},
  pdfpagelayout = {SinglePage},
  pdfstartview = {Fit},
  pdftoolbar = {true},
  plainpages = {false},
}

\usepackage[utf8x]{inputenc} 
\usepackage{lipsum}

\allowdisplaybreaks

\newcommand{\Fref}[1]{Fig.~\ref{#1}}
\newcommand{\Fsref}[1]{Figs.~\ref{#1}}

\newcommand{\Eqref}[1]{Eq.~(\ref{#1})}

\newcommand{\Tbref}[1]{Tab.~\ref{#1}}

\newcommand{\Acal}{\mathcal{A}}
\newcommand{\Ecal}{\mathcal{E}}
\newcommand{\Fcal}{\mathcal{F}}
\newcommand{\Jcal}{\mathcal{J}}

\newcommand{\bk}{\boldsymbol{\kappa}}

\usepackage[usenames,dvipsnames]{xcolor}

\graphicspath{ {./Figs/} } 

\begin{document}

\title{Iterative approach to arbitrary nonlinear optical response functions 
of graphene}

\author{F. Hipolito}

\email{fh@nano.aau.dk}

\affiliation{Department of Physics and Nanotechnology,
Aalborg University, DK-9220 Aalborg {\O}st, Denmark}

\author{Darko Dimitrovski}

\affiliation{Department of Physics and Nanotechnology,
Aalborg University, DK-9220 Aalborg {\O}st, Denmark}
%
%

\author{T. G. Pedersen}
\email{tgp@nano.aau.dk}
\affiliation{Department of Physics and Nanotechnology,
Aalborg University, DK-9220 Aalborg {\O}st, Denmark}
\affiliation{Center for Nanostructured Graphene (CNG), 
DK-9220 Aalborg {\O}st, Denmark}

\begin{abstract}
Two-dimensional materials constitute an exciting platform for nonlinear optics  
with large nonlinearities that are tunable by gating.
Hence, gate-tunable harmonic generation and intensity-dependent refraction have 
been observed in e.g. graphene and transition-metal dichalcogenides, whose 
electronic structures are accurately modelled by the (massive) Dirac equation.
We exploit on the simplicity of this model and demonstrate here that 
arbitrary nonlinear response functions follow from a simple iterative approach. 
The power of this approach is illustrated by analytical expressions for 
harmonic generation and intensity-dependent refraction, both computed up to 
ninth order in the pump field.
Moreover, the results allow for arbitrary band gaps and gating potentials.
As illustrative applications, we consider 
(i) gate-dependence of third- and fifth-harmonic generation in gapped and 
gapless graphene, 
(ii) intensity-dependent refractive index of graphene up to ninth order, and 
(iii) intensity-dependence of high-harmonic generation. 
\end{abstract}

\pacs{42.65.An,78.67.-n,78.67.Wj,81.05.ue}


\maketitle
%
%

Nonlinear optical (NLO) response encompasses a large class of light matter 
interactions \cite{Boyd2008, Shen2002, Franken1963, Bloembergen1982, Axt1998, 
Kuzyk2013}, including processes such as harmonic generation and 
self-focusing of light, that have proven useful in a number of 
applications in nonlinear spectroscopy and in optoelectronic devices.
Recent progress in the fabrication of 2D materials has produced a new fertile 
class of materials with large nonlinear susceptibilities \cite{Autere2018}.
Recent reports include measurements of high harmonic generation (HHG) 
\cite{Yoshikawa2017, Hafez2018} and intensity dependent refractive index 
\cite{Lim2011, Mohsin2015} in graphene and in transition metal dichalcogenides 
(TMDs) \cite{Liu2017}.
In addition, it has been shown that the NLO response can be tuned 
by electrostatic doping \cite{Soavi2017, Jiang2018, Zhang2018a} and significant 
progress has been made in measuring even-order NLO response in TMDs 
\cite{Wang2015a, Saynatjoki2017}.
Furthermore, the nonlinearities in 2D materials can be significantly enhanced 
by several mechanisms such as plasmons \cite{Cox2017, Kundys2017, Wang2018},
polaritonic effects \cite{Wild2018}, and metasurfaces \cite{Rosolen2018}.

Compared with the linear response, calculations of NLO processes in crystals
are significantly more complex.
Whereas the linear response results from purely inter- or intraband processes, 
the NLO response contains not only these processes, but also mixed ones 
involving both inter- and intraband motion of electrons \cite{Aversa1995, 
Pedersen2015, Taghizadeh2017a, Hipolito2018a}.
To circumvent this complexity, the NLO response has been characterized using 
several theoretical methods, each with its own merits and shortcomings:
(i) perturbative expansion of the reduced density matrix \cite{Cheng2014, 
Rostami2016, Hipolito2018a};
(ii) time-dependent techniques \cite{Tamaya2016, Chizhova2017, Dimitrovski2017, 
Mikhailov2017};
(iii) Wannier representation \cite{Catoire2018}.
The perturbative method offers a feasible approach to specific processes at a 
fixed frequency and power of the external field.
The standard approach expands all matrix elements in unperturbed eigenstates, 
leading to increasingly complicated sum-over-states expressions for high-order 
processes.
Still, within the perturbative regime, highly accurate results are obtained and 
in simple few-band systems such as the Dirac Hamiltonian, closed form solutions 
can often be found. These allow for characterization with respect to external 
parameters, for instance doping and temperature \cite{Cheng2014, Rostami2016, 
Hipolito2018a}.
Yet, the growth in complexity associated with mixed inter- and intraband motion 
makes calculations extremely cumbersome beyond third order. 
The complicated nature of the general third-order response \cite{Hipolito2018a}
testifies to this complexity.
Methods (ii) and (iii) can be applied to study the NLO response at field 
strengths beyond the perturbative regime, as these intrinsically include 
contributions from all powers of the external field.
But, in contrast to perturbative approaches, these methods rely extensively on 
numerical techniques for the integration of the equation of motion and for the 
Fourier transforms required to analyze the response in frequency domain, thus 
making the characterization of the NLO response with respect to external 
parameters an elaborate numerical process \cite{Tamaya2016, Chizhova2017, 
Dimitrovski2017, Mikhailov2017}.

\begin{figure}
\includegraphics[width=1.00\linewidth]{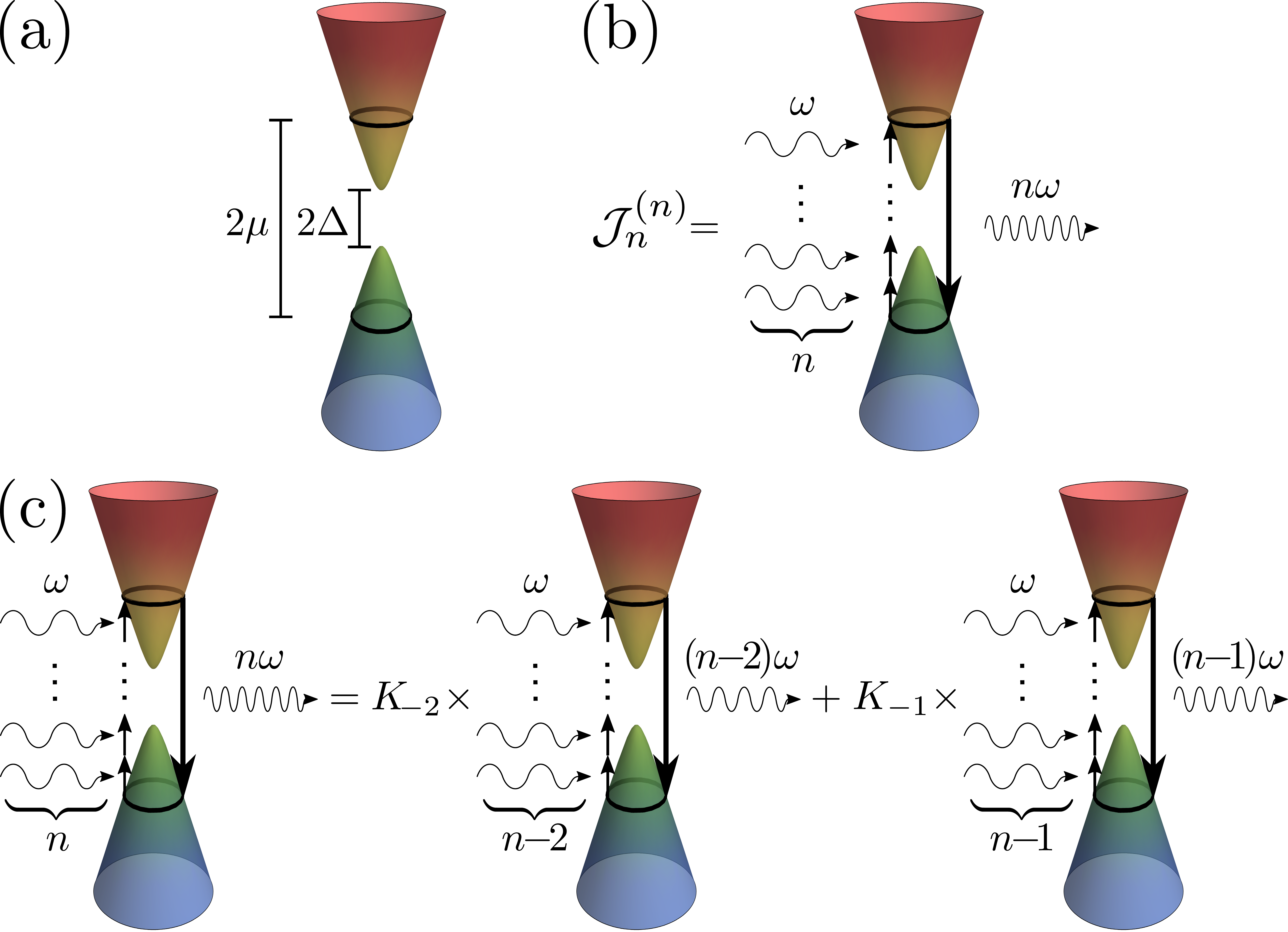}
\caption[Iterative solution]{\label{fig:diag}%
Diagrammatic representation of the band structure (a), of the $n^\mathrm{th}$ 
harmonic integrand $\Jcal _n ^{(n)}$ at the resonance $2\mu = n \omega$ (b) and 
of the iterative solution for $\Jcal _n ^{(n)}$ integrand (for $n^\mathrm{th}$ 
harmonic) as in \Eqref{eq:J:harmonic}.
Here, $\Jcal _n ^{(n)}$ represents absorption of $n$ 
photons with frequency $\omega$, followed by the emission of a single 
photon with frequency $n \omega$.}
\end{figure}

In the present letter, we study the (massive and massless) Dirac Hamiltonian as
a model of graphene and TMDs.
For this important class of materials, we bridge a key shortcoming found in 
perturbative techniques by evaluating the current density response via an 
iterative solution.
This approach allows for the evaluation of arbitrarily high order response 
functions.
As an illustration, we compute all response functions up to ninth order for 
(gapped) graphene 
\footnote{See Supplemental Material at [URL will be inserted by publisher] for 
the full expressions for gapped graphene and for additional information.}.
The massive Dirac Hamiltonian \cite{Semenoff1984} with a external vector
potential $\mathbf{A} = \mathbf{A}_0 \sin(\omega t) $ reads, using the minimal 
coupling (velocity gauge), \cite{Peres2010}
\begin{equation}
\label{eq:H}
H = v_F ( \hbar \boldsymbol{\kappa} +e \mathbf{A} )\cdot \boldsymbol{\sigma}
+\hbar \Delta \sigma_z ,
\end{equation}
where $v_F \sim 10^6 \ \mathrm{m/s}$ is the Fermi velocity, $\bk$ is the
wavevector, $\boldsymbol{\sigma}$ are the Pauli matrices, $\Delta \geq 0$ is 
the mass term, and $e>0$.
This model leads to a gapped band structure with energy gap
$E_g = 2\hbar\Delta$, and doping is including via a non-vanishing Fermi level 
$\hbar\mu$. Hence, for pristine graphene, we take $\Delta=0$.
The key features of the electronic structure are shown in \Fref{fig:diag}(a).
The Dirac Hamiltonian has proven extremely useful for systems with threefold 
rotation symmetry.
It allows for accurate analytic characterization of several physical properties 
in graphene \cite{CastroNeto2009} and in the vast class of TMDs 
\cite{Xiao2012}.

The time evolution of the wave function $\psi$ governed by
$ i\hbar \dot{\psi} = H \psi$ is found by expanding in the eigenstates 
$u_{1,2}$ of the unperturbed Hamiltonian, i.e. taking $\mathbf{A} =0$.
We write the general wave function as
\begin{equation}
\label{eq:psi:general}
\psi = [ a(t) u_1 +b(t) u_2 ] 
\exp[ i(\boldsymbol{\kappa} \cdot \mathbf{r} +\epsilon t) ] ,
\end{equation}
with energy dispersion $ \epsilon = \sqrt{ \Delta^2 +k^2 }$ (in frequency
units), and $k = v_F |\bk|$.
Furthermore, we focus on the response to a normally incident, 
linearly polarized monochromatic field, $\mathbf{A} = A \mathbf{e}_x$ and 
define 
$\Acal(t) = v_F e A_0\sin(\omega t) /\hbar = \Acal_0 \sin(\omega t)$,
which is related to the electric field $\Ecal(t) = \Ecal_0\cos(\omega t)$ by 
$\Ecal_0 = -\omega A_0$.
The time evolution of the coefficients follows from
$\dot{a}(t) = i \Acal(t) [ a(t) F -b(t)G ]$
and 
$i\dot{b}(t) = \Acal(t)  G^* a(t) +[ 2\epsilon +\Acal(t) F ] b(t)$, where 
$F = \sqrt{\epsilon^2-\Delta^2} \cos( \theta )/ \epsilon $ and
$G = [ \Delta \cos(\theta) -i\epsilon \sin(\theta) ]/ \epsilon $ arise from
matrix elements of the velocity operator.
The coefficients then determine the reduced density matrix, whose 
matrix elements read
$\rho_{11} = |a|^2$, $\rho_{22} = |b|^2$ and $\rho_{21} = a^* b$.
In turn, their time evolution is governed by
\begin{subequations}
\begin{align}
i \dot{\mathcal{N}} &= -2 \Acal(t) ( \mathcal{P} G - G^* \mathcal{P}^*  ) ,
\\
i\dot{\mathcal{P}} &= 
-\Acal(t) G^* \mathcal{N} +2\big[ \Acal(t) F +\epsilon ] \mathcal{P},
\end{align}
\end{subequations}
where $\mathcal{N} = \rho_{22} -\rho_{11}$ and $\mathcal{P}=\rho_{21}$ define 
the population difference and coherence, respectively.
Finally, the current density is evaluated via the expectation value of the 
current density operator
$j = -e g_s g_v (2\pi)^{-2} v_F \int \Jcal d \boldsymbol{\kappa} $, 
where the dimensionless integrand for the current density response is defined 
by $\Jcal = \sum_{mn} v_{nm} \rho_{mn}/v_F
= F \mathcal{N} +\mathcal{P}G +  G^*\mathcal{P}^*$
using the matrix elements $v_{nm}$ of the velocity operator 
$\hat{v}_x = v_F \hat{\sigma}_x$. Here, $g_s=2$ and $g_v=2$ are spin and valley
degeneracies, respectively.

The iterative sequence is found by considering the first- and second-order time
derivatives of $\Jcal$ that read
\begin{subequations}
\begin{align}
\Acal(t) \dot{\Jcal} &= -\epsilon \dot{\mathcal{N}},
\\
\ddot{\Jcal} &= 4\epsilon \big[ \Acal(t) +\epsilon F \big]\mathcal{N}
-4\epsilon \big[ \Acal(t) F +\epsilon \big] \Jcal.
\end{align}
\end{subequations}
Using a time-harmonic expansion for the integrand
$\Jcal = \sum_n \Jcal_n e^{-i n \omega t }$ and for the population
$\mathcal{N} = \sum_n \mathcal{N}_n e^{-i n \omega t }$, the dynamical 
equations can be cast as
\begin{subequations}
\begin{align}
  \label{eq:J1}
  \Acal_0[ (n-1) \Jcal_{n-1} -(n+1)\Jcal_{n+1} ]&= 2i n \epsilon \mathcal{N}_n ,
\end{align}
\begin{align}
  \label{eq:J2}
(4\epsilon^2 -n^2 \omega^2 ) \Jcal_n &=
 4\epsilon^2 F \mathcal{N}_n
-2i\epsilon \Acal_0 F( \Jcal_{n-1} -\Jcal_{n+1} )
\nonumber\\&
 +2i \epsilon  \Acal_0 ( \mathcal{N}_{n-1} -\mathcal{N}_{n+1} ) ,
\end{align}
\end{subequations}
where $n\in \mathbb{Z}$ defines the Fourier order.
The final iterative series for the integrand is identified by making use of an 
expansion with respect to powers of the external field
$\Jcal_n = \sum_{j\geq n} \Jcal_n^{(j)} \Acal_0^j$ and collecting equal powers
\footnote{Note:~it is sufficient to consider $n\geq 0$, as the terms for 
$\Jcal_{-n}^{(j)}$ can be immediately obtained from $\Jcal_{n}^{(j)}$ by means 
of the replacement $\omega \to -\omega$.}%
\begin{align}
\label{eq:Jiterative}
\Jcal_n^{(j)} &=
 \Theta_{j,n+4} K_{2} \Jcal_{n+2}^{(j-2)}
\nonumber\\&
+\Theta_{j,n+2} \Big[
   \Theta_{j,2} \bar\delta_{n,1} K_{ 0} \Jcal_{n  }^{(j-2)} 
  +             \bar\delta_{n,0} K_{ 1} \Jcal_{n+1}^{(j-1)} \Big]
\nonumber\\&
+\Theta_{j,n} \Big[  
   \Theta_{j,2}     K_{-2} \Jcal_{n-2} ^{(j-2)}
  +\bar\delta_{n,0} K_{-1} \Jcal_{n-1} ^{(j-1)} \Big] ,
\end{align}
where $n+j \geq 1$, 
$\Theta_{i,j}$ is the discrete unit step function %
\footnote{Note:~the discrete unit step function is defined as
$\Theta_{i,j} = 1 \, \mathrm{for} \, i \geq j$, 
$\Theta_{i,j} = 0 \, \mathrm{for} \, i < j$.}, 
$\bar\delta_{i,j} \equiv 1 -\delta_{i,j}$ and the coefficients read:
$K_{0} = 2n^2/[Q_n(n-1)]$; 
$K_{\pm1} = \mp 2iF \epsilon(2n \pm 1)/(Q_n n)$ and
$K_{\pm2} = -(n \pm 2)/[Q_n(n \pm 1)]$ with
$Q_n = 4\epsilon^2 -n^2 \omega^2$.
The dominant term in harmonic generation emerges from the diagonal case 
$j=n>2$, where 
the general solution \Eqref{eq:Jiterative} reduces to
\begin{equation}
\label{eq:J:harmonic}
\Jcal_n^{(n)} =
  K_{-2} \Jcal_{n-2} ^{(n-2)}
+ K_{-1} \Jcal_{n-1} ^{(n-1)} ,
\end{equation}
which lends itself to a diagrammatic representation as illustrated in  
\Fref{fig:diag}(b-c).
%
%
%
%
%

To apply the iterative solution for $\Jcal_n^{(j)}$ in practice, two seeds
$\Jcal_0^{(0)}$ and $\Jcal_1^{(1)}$ are required that can easily be 
determined from low-order terms in \Eqref{eq:J2}.
Collecting all terms independent of the external field and making use of the 
equilibrium charge distribution $\mathcal{N}_0 = f_1 -f_2$ (the difference 
between Fermi functions), the first seed reads 
$\Jcal_0^{(0)} = F \mathcal{N}_0$.
The second seed involves the collection of linear terms in the external field
and reads $\Jcal_1^{(1)} = 2i \epsilon |G|^2 \mathcal{N}_0 \Acal_0
/[ 4\epsilon^2 -\omega^2]$.
All remaining terms of $\Jcal_n^{(j)}$ can be computed sequentially by 
evaluating all possible Fourier components $n=\{0,1,\ldots,j\}$, in increasing 
order, for any given response order $j$ using \Eqref{eq:Jiterative}.
The solutions for all nonzero integrands up to fifth order are listed in this 
order in \Tbref{tab:J}.
%
%
%
\begin{table*}[t]
\caption[Current density integrands]{\label{tab:J}%
Analytic expressions for current density 
integrands up to fifth order, where 
$\Jcal _n^{(j)} \equiv \Fcal_n^{(j)} |G|^2 \mathcal{N}_0\Acal_0^j $.}
\begin{ruledtabular}
\begin{tabular}{ll}
$(n,j)$ & $\Fcal_n^{(j)} $\\ \hline
$(0,0)$ & $ F/|G|^2$\\
\hline
$(1,1)$ & $ 2i\epsilon
    \big/[ ( 4\epsilon^2 -\omega^2 ) ]$\\
\hline
$(0,2)$ & $-2 F
       \big/[ 4\epsilon^2 -\omega^2 ]$\\
$(2,2)$ & $ 6 \epsilon^2 F
    \big/ \big[ (4\epsilon^2 -4\omega^2)(4\epsilon^2 -\omega^2) \big]$\\
\hline
$(1,3)$ & $ 4i\epsilon
    \big[ (13F^2-1)\epsilon^2 -(4F^2-1)\omega^2 \big]
    \big/ \big[ (4\epsilon^2 -4\omega^2) (4\epsilon^2 -\omega^2)^2 \big]$\\
$(3,3)$ & $-4i \epsilon \big[ (5F^2-1)\epsilon^2 +\omega^2 \big] 
    \big/ \big[ 
    (4\epsilon^2 -9\omega^2)
    (4\epsilon^2 -4\omega^2)
    (4\epsilon^2 - \omega^2) \big]$\\
\hline
$(0,4)$ & $-2 F\big[ 2(13F^2-7)\epsilon^2 -(8F^2-5)\omega^2 \big]
    \big/
    \big[ (4\epsilon^2 -4\omega^2) (4\epsilon^2 -\omega^2)^2 \big] $\\
$(2,4)$ & $ 16\epsilon^2 F \big[ 8(8F^2-3)\epsilon^4 
    -2(53F^2 -28)\epsilon^2\omega^2 +(27F^2-17)\omega^4  \big]
    \big/ \big[ 
    (4\epsilon^2 -9\omega^2)
    (4\epsilon^2 -4\omega^2)^2
    (4\epsilon^2 - \omega^2)^2 \big]$\\
$(4,4)$  & $-10\epsilon^2 F \big[ (7F^2-3)\epsilon^2 +5\omega^2 \big] 
    \big/ \big[ 8
    (4\epsilon^2 -16\omega^2)
    (4\epsilon^2 - 9\omega^2)
    (4\epsilon^2 - 4\omega^2)
    (4\epsilon^2 -  \omega^2)\big] $\\
\hline
$(1,5)$ & $ 48i \epsilon \big[ 
     8( 61F^4 -32F^2 +1)\epsilon^6 
    -2(503F^4-334F^2+11)\epsilon^4 \omega^2
    +5(100F^4 -77F^2 +4)\epsilon^2 \omega^4$ \\
    & 
    $ 
    -3( 24F^4 -21F^2 +2) \omega^6 \big] 
    \big/ \big[ 
    (4\epsilon^2 -9\omega^2)
    (4\epsilon^2 -4\omega^2)^2
    (4\epsilon^2 - \omega^2)^3 \big] $\\
$(3,5)$ & $-16i \epsilon \big[ 
      4( 295F^4 -186F^2 +11)\epsilon^8
    -  (5235F^4-4166F^2+291)\epsilon^6 \omega^2
    +  (6095F^4-6303F^2+558)\epsilon^4 \omega^4
    -  (1440F^4 $ \\
    & 
    $
    -2569F^2+419)\epsilon^2 \omega^6
    +36(3-8F^2)\omega^8 \big] 
    \big/ \big[ 
    (4\epsilon^2 -16\omega^2)  
    (4\epsilon^2 - 4\omega^2)^2
    (4\epsilon^2 - 9\omega^2)^2
    (4\epsilon^2 -  \omega^2)^2 \big] $\\
$(5,5)$ & $ 12 i\epsilon \big[ 
      ( 21F^4 -14F^2 +1)\epsilon^4
    -5( 1 -7F^2)\epsilon^2 \omega^2
    +4\omega^4 \big]
    \big/ \big[
    (4\epsilon^2 -25\omega^2)
    (4\epsilon^2 -16\omega^2)
    (4\epsilon^2 - 9\omega^2)
    (4\epsilon^2 - 4\omega^2)(4\epsilon^2 -  \omega^2) \big]$\\
\end{tabular}
\end{ruledtabular}
\end{table*}
%
%
%
The final response functions are obtained by integrating the desired 
$\Jcal_n^{(j)}$ over $\kappa$-space and the respective conductivities
$\sigma_n^{(j)}( \omega )$ then follow by writing
\begin{equation}
j(t) = \sum_{n,j} 
\big[ \sigma_n^{(j)}( \omega ) \Ecal_0^j e^{-i n \omega t } 
+\mathrm{c.c.} \big]/2.
\end{equation}
In most cases, the integration is straightforward, but can lead to cumbersome 
expressions, particularly whenever the difference between the Fourier 
$n$ and response order $j$ is large.
The angular part of the integral depends exclusively on powers of $F$ and 
$|G|$, therefore it can be shown that due to the presence of full rotation 
symmetry in the effective Hamiltonian all even-order response functions vanish 
upon angular integration.
Nonetheless, even-order integrands remains necessary to 
determine higher order non-vanishing odd integrands.

Now, we turn our attention to the conductivities computed within 
the iterative framework.
At low temperatures the population difference becomes a step function
$\mathcal{N}_0 = -\Theta ( \epsilon -|\mu| )$ and the lower limit $\Omega$ of
the radial part of the integral is determined by the larger of the Fermi level 
and mass term, i.e. $\Omega \equiv \max( |\mu|, \Delta )$
\footnote{Note:~the integral over the wavevector has been replaced by an 
integration over energy.}.
In our explicit examples, we compute all conductivities up to ninth order 
\cite{Note1}.
Among these, we examine in detail third and fifth harmonic generation as well 
as intensity-dependent refraction through the optical Kerr effect including 
high-order terms.
As demonstrated in recent experiments \cite{Soavi2017, Jiang2018, Zhang2018a}, 
valuable information can be extracted by varying the Fermi level via 
electrostatic gating.
Hence, we apply the present results to study the doping dependence of these NLO 
processes.

\begin{figure}
\includegraphics[width=1.00\linewidth]{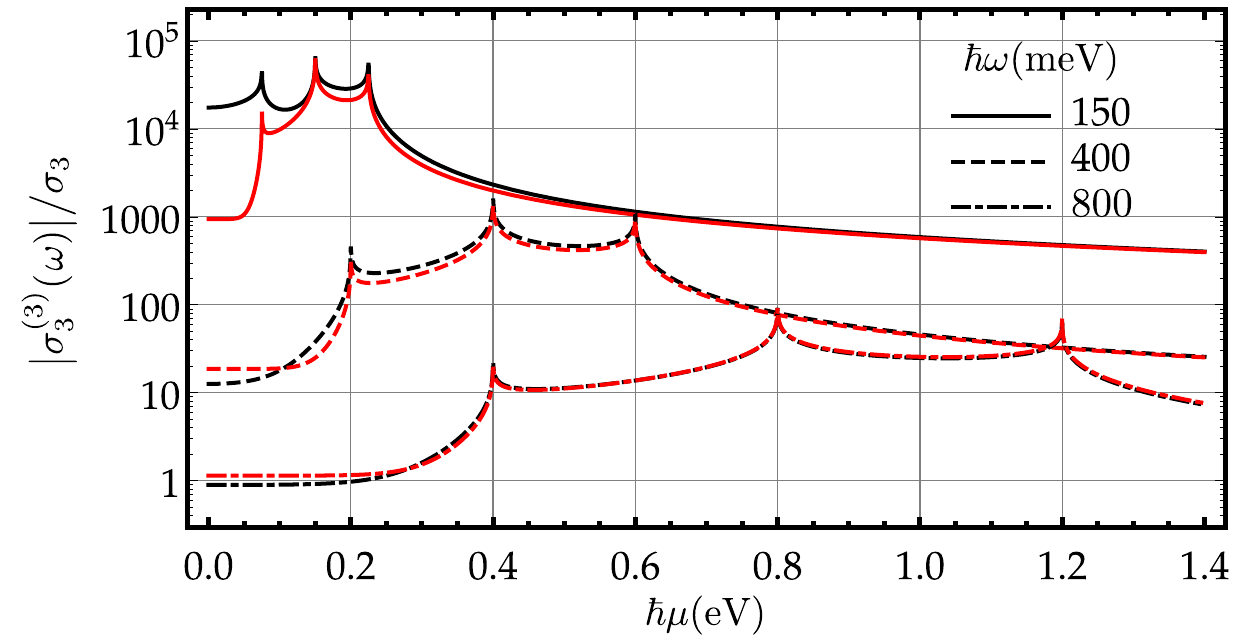}
\caption[THG in gapped graphene]{\label{fig:3HG}%
THG conductivity in gapped ($E_g =250$ meV, black) and gapless (red) 
graphene at photon energies $\hbar\omega = \{ 150, 400, 800 \}$ 
meV.}
\end{figure}

The third harmonic generation (THG) conductivity reads
\begin{align}
\label{eq:gapped:THG}
 \sigma_3^{(3)}(\omega) &= 
 \frac{ -3i\sigma_3 }{ 1024 \pi }
 \bigg( \frac{ 2 v_F }{ 3 a_0 \omega } \bigg)^4
\nonumber\\&
\bigg[ \bigg( 45
+\frac{ 56 \Delta^2 }{ \omega^2 }
-\frac{ 48 \Delta^4 }{ \omega^4} \bigg)
 \ln \frac{ 2\Omega -3\omega}{ 2\Omega +3\omega }
\nonumber\\&
-\bigg( 64 
  +\frac{ 128\Delta^2 }{ \omega^2 } 
  -\frac{ 192\Delta^4 }{ \omega^4 } \bigg) 
 \ln \frac{ 2\Omega -2\omega }{ 2\Omega +2\omega } 
\nonumber\\&
+\bigg( 17
+\frac{ 88 \Delta^2  }{ \omega^2 }
-\frac{ 240 \Delta^4 }{ \omega^4 } \bigg)
 \ln \frac{ 2\Omega - \omega }{ 2\Omega + \omega }
\bigg] ,
\end{align}
where we define the scale of the 2D nonlinear conductivities systematically by 
$\sigma_{j>1} \equiv 2[ 3 e a_0^2 /(4 \hbar v_F )]^{j-1} \sigma_1$ with
$\sigma_1 = e^2/ (4\hbar)$ and the carbon-carbon distance $a_0 \equiv 1.42$
\AA\ sets the natural length scale for graphene.
Throughout the letter, we consider exclusively electron doping $\mu > 0$, but
results for hole doping $\mu<0$ simply follow by replacing $\mu \to -\mu$.
Taking the limit $\Delta \to 0$, one can verify that our expression 
reduces to previous results derived with the gapless Dirac Hamiltonian using 
velocity and length gauges \cite{Cheng2014, Rostami2016, Hipolito2018a}.
The expression for $\sigma_3^{(3)}(\omega)$ is representative of the HHG 
conductivities, $\sigma_n^{(n)}(\omega)$, which are always composed by $n$ 
logarithmic divergences, whose amplitude is set by a polynomial prefactor with 
even powers of $\Delta/\omega$ as shown in Eq.~S1a (see supplemental material).
Note that the divergences found in the expressions are regularized by 
introducing a small broadening parameter $\omega \to \omega +i \eta$ and, 
unless stated otherwise, we use $\hbar\eta = 1$ meV.
The THG conductivities for both gapped and gapless graphene assuming photon
energies in the low and medium range are shown in \Fref{fig:3HG}.
Given the rather small gaps that can be reliably generated in graphene 
\cite{Zhou2007, Woods2014} (we take $E_g \equiv 250$ meV as a reference figure 
for our calculations) and considering photon energies $\hbar\omega > 150$ meV, 
our results show that the response of gapless and gapped systems are generally
similar but deviate whenever $\mu \lesssim \Delta$.
%
%
%
%
%

\begin{figure}
\includegraphics[width=1.00\linewidth]{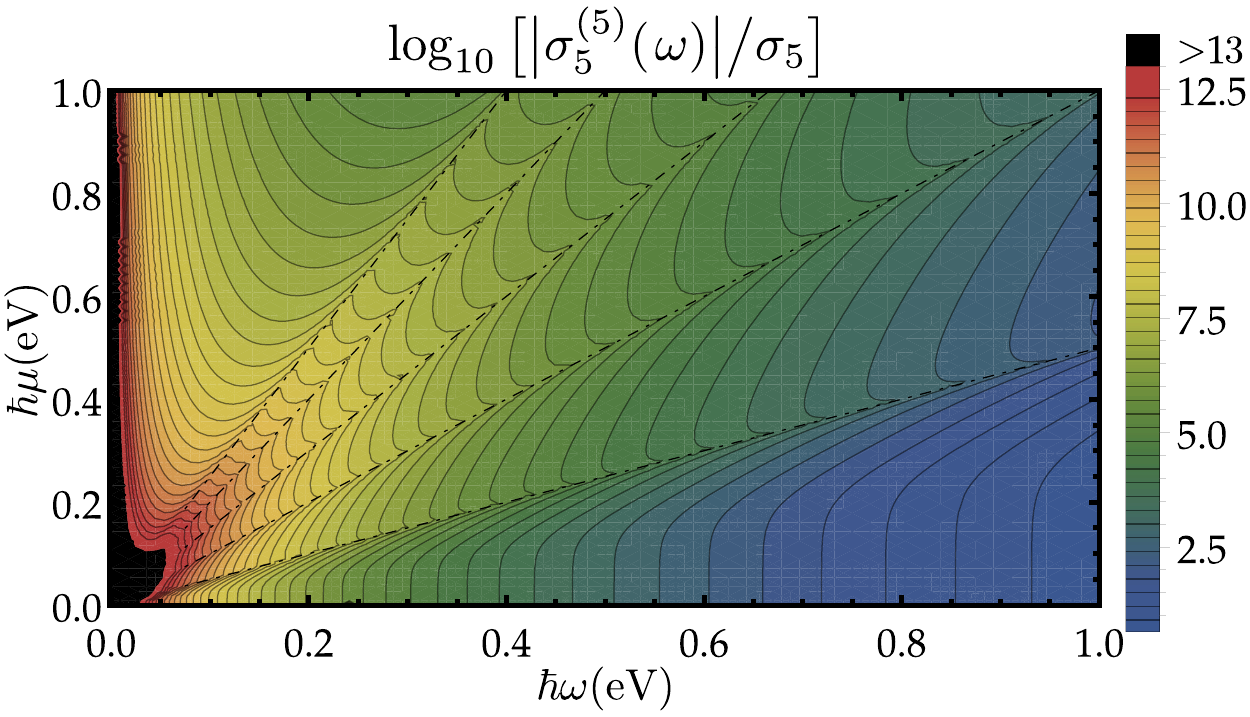}
\caption[Fifth harmonic in graphene]{\label{fig:5HG}%
Contour plot of fifth harmonic conductivity in graphene. Dot-dashed lines show 
the resonant conditions $\mu =m \omega/2 \,, m = \{1,2,3,4,5\}$ and the 
conductivity in the black region exceeds $10^{13} \sigma_5$.}
\end{figure}

The fifth harmonic conductivity in gapless graphene also lends itself to a 
closed form expression \cite{Note1}, with $n=5$ logarithmic divergences
\begin{align}
\label{eq:graph:FHG}
\sigma_{5}^{(5)}(\omega) &= 
\frac{ 9i \sigma_5 }{ 655360 \pi } 
 \bigg( \frac{ 2 v_F }{ 3 a_0 \omega } \bigg)^8
\bigg[
  4925 \ln \frac{2\mu-5\omega}{2\mu+5\omega}
\nonumber\\&
-16384 \ln \frac{2\mu-4\omega}{2\mu+4\omega}
+19359 \ln \frac{2\mu-3\omega}{2\mu+3\omega}
\nonumber\\&
- 9216 \ln \frac{2\mu-2\omega}{2\mu+2\omega}
+ 1266 \ln \frac{2\mu- \omega}{2\mu+ \omega}
\bigg] ,
\end{align}
where the general expression valid for $\Delta \neq 0$ is given in Eq.\ (S3)
\cite{Note1}.
The fifth-order response of graphene is highly sensitive to the ratio 
between photon energy and doping level.
This is illustrated in the contour plot in \Fref{fig:5HG} of the 
amplitude of the fifth harmonic conductivity as function of these parameters.
It shows that this response function can be tuned over several orders of 
magnitude by tuning either parameter, while highlighting the five resonances 
present in the fifth harmonic response.
Moreover, we find that the nonlinear conductivities of graphene are regular in 
the limit of vanishing doping $\mu \to 0$
\begin{align}
\lim_{\mu \to 0} \sigma_n^{(j)} / \sigma_j  = 
q_{nj} \big[  2 v_F /( 3a_0 \omega ) \big]^{2(j-1)} \, ,
\end{align}
where the coefficients $q_{nj}$ are rational numbers.
The complete list for all coefficients is found in Tab. S1 in supplemental 
material. For third and fifth harmonic generation, the coefficients read 
$q_{33} = 3/512$ and $q_{55} = -45/65536$, respectively.
%
%
%
%
%
%

The iterative approach can also readily be used to evaluate conductivities
beyond harmonic generation such as the optical Kerr conductivity of graphene 
\cite{Note1}
\begin{align}
\label{eq:sigma:Kerr}
\sigma_{1}^{(3)}( \omega) &= 
\frac{ 9i\sigma_3 }{ 256 \pi }
 \bigg( \frac{ 2 v_F }{ 3 a_0 \omega } \bigg)^4
\bigg[
\frac{ 12\mu\omega }{ 4\mu^2 -\omega^2 }
\nonumber\\&
-11 \ln \frac{ \omega -2\mu}{ \omega +2\mu}
+16 \ln \frac{2\omega -2\mu}{2\omega +2\mu}
\bigg] .
\end{align}
This expression is representative of high-order contributions to any 
Fourier order $\sigma_n^{(j>n)}$ \cite{Note1}.
These expressions contain $(j+n)/2$ logarithmic divergences, rather than $j=n$ 
found in the $n^\mathrm{th}$ harmonic conductivities, and also contain an 
additional rational function with $(j+n-2)/2$ polynomial divergences that 
strongly enhance the nonlinear resonances, see Fig.\ S1 in supplemental 
material.
In \Fref{fig:Kerr}, we show the conductivities $\sigma_1^{(j)}$ contributing to 
the first harmonic current in doped graphene up to ninth order at
$\Ecal_0 = 2 \, \mathrm{V/ \mu m}$.
Note that the field intensity considered in \Fref{fig:Kerr} matches the upper 
limit of the perturbative regime when considering THz radiation 
\cite{Hafez2018}.
Inspection of \Fref{fig:Kerr} defines the regime, where the perturbative 
approach breaks down, namely the frequency range, in which terms 
$\sigma_1^{(j)} \Ecal _0 ^{j-1}$ cease to decrease as the order $j$ is 
increased.
Hence, for the parameters in \Fref{fig:Kerr} the non-perturbative region can be 
estimated as $\hbar\omega \lesssim 60$ meV.
Manifestations from higher than Kerr terms should be detectable as higher order 
terms introduce additional resonances that are highly sensitive to both the 
Fermi level and the magnitude of the external field.
%
%
%
%
\begin{figure}
\includegraphics[width=1.00\linewidth]{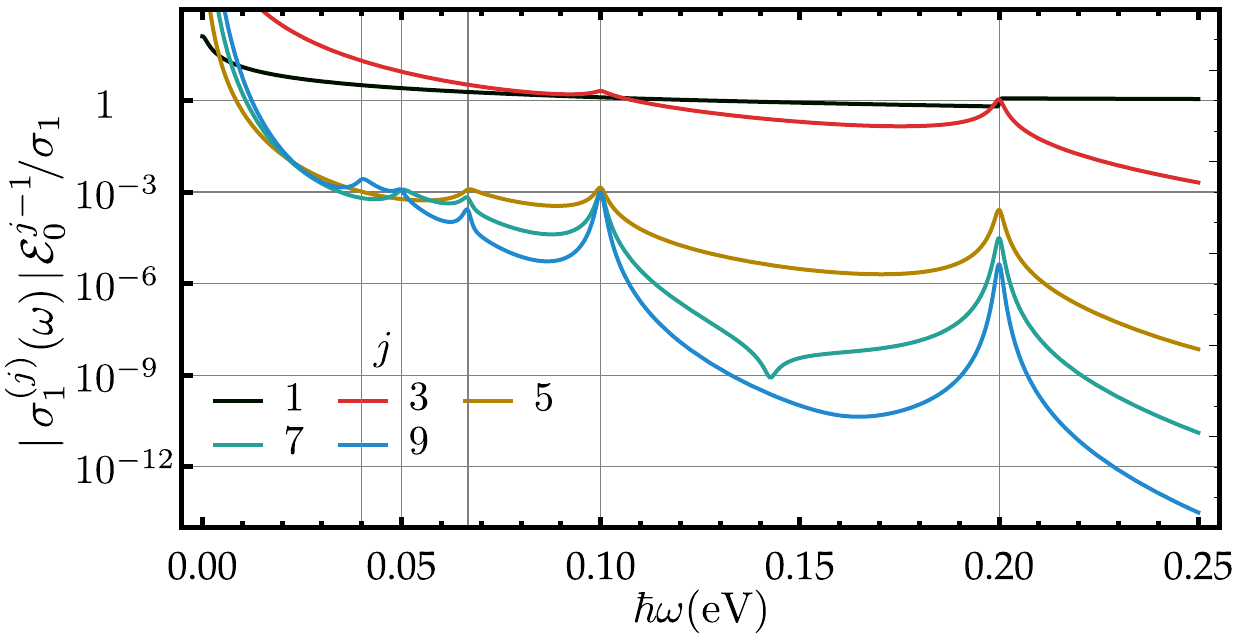}
\caption[Optical conductivity in graphene]{\label{fig:Kerr}%
Linear and nonlinear contributions to the optical conductivity in doped 
graphene 
$\hbar\mu = 100 \, \mathrm{meV}$ at $\Ecal_ 0 = 2 \, \mathrm{V/ \mu m}$.
The black curve is the linear response and colored lines represent the 
nonlinearities.
Vertical grid lines represent the $(n+j)/2$ resonances found in the nonlinear 
conductivities.}
\end{figure}%
%
%
%

\begin{figure}
\includegraphics[width=1.00\linewidth]{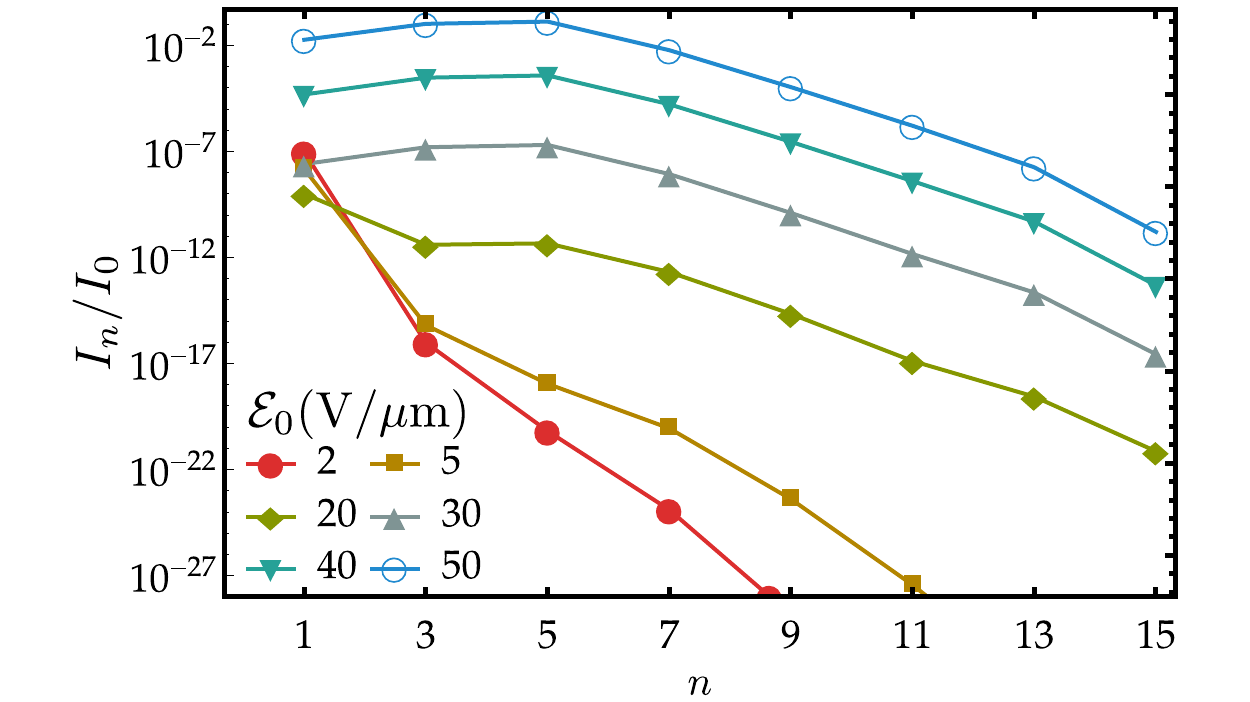}
\caption[Fourier components of current response]{\label{fig:nHG}%
Intensity of HHG Fourier components in doped graphene ($\hbar\mu = 250$ meV) 
normalized by $I_0 = \varepsilon_0 c_0 \Ecal_0^2/2$.
The incident photon energy is $\hbar\omega = 100$ meV and colors red to blue 
represent increasing field strengths.}
\end{figure}

In \Fref{fig:nHG}, we plot the relative amplitude of the Fourier components of
the radiated intensity $I_n(\omega) = \mu_0 c |j_n(\omega)|^2/8$  with 
$j_n(\omega) = \sum_j \sigma_n^{(j)}(\omega) \Ecal_0^j$ with respect to the 
incident intensity in vacuum $I_0 = \varepsilon_0 c_0 \Ecal_0^2/2$ 
\cite{Stauber2008,Hipolito2017} considering all contributions up to $j=15$.
Note that the analytic expressions for the conductivities are limited to 
ninth order, hence all data plotted in \Fref{fig:nHG} were integrated 
numerically using $\hbar\eta = 10$ meV. 
Results shown in \Fsref{fig:3HG} to \ref{fig:nHG} demonstrate that the approach 
presented in this letter can be used to readily characterize harmonic response 
of graphene, including the effects from higher order terms, at arbitrary doping 
level and photon frequency, without requiring the complex numerical 
calculations found in time-dependent techniques. 
%
%
%
%

%
%
%

In summary, we introduce and apply an iterative approach to the calculation of
NLO response of systems described by the massive Dirac Hamiltonian.
The iterative nature allows for analytical evaluation of high-order response
functions, and we derive for the first time all nonlinear conductivities of 
(gapped) graphene up to ninth order.
The NLO response of doped graphene reveals an intricate interplay between 
doping, photon energy and the intensity of the external electric field.

%
%
%

\begin{acknowledgments}
The authors acknowledge Alireza Taghizadeh for many helpful comments.
This work was supported by the QUSCOPE center sponsored by the Villum 
Foundation, and TGP is supported by the CNG center under the Danish National 
Research Foundation, project DNRF103. 
\end{acknowledgments}


\bibliography{iterative}


\end{document}


\title{Supplemental Material for: Iterative approach to arbitrary nonlinear 
optical response functions of graphene}

\author{F. Hipolito}

\email{fh@nano.aau.dk}

\affiliation{Department of Physics and Nanotechnology,
Aalborg University, DK-9220 Aalborg {\O}st, Denmark}

\author{Darko Dimitrovski}

\affiliation{Department of Physics and Nanotechnology,
Aalborg University, DK-9220 Aalborg {\O}st, Denmark}
%
%

\author{T. G. Pedersen}
\email{tgp@nano.aau.dk}
\affiliation{Department of Physics and Nanotechnology,
Aalborg University, DK-9220 Aalborg {\O}st, Denmark}
\affiliation{Center for Nanostructured Graphene (CNG), 
DK-9220 Aalborg {\O}st, Denmark}

\begin{abstract}
We provide supplemental material for:
i) the role of high order contributions to the Kerr conductivity;
ii) the impact of doping on different Fourier components of the radiated field;
iii) expressions for all nonlinear conductivities in (gapped) graphene
up to ninth order.
\end{abstract}

\pacs{42.65.An,78.67.-n,78.67.Wj,81.05.ue}


\maketitle

\section{Interplay between logarithmic and polynomial parts}
\begin{figure}[b]
\includegraphics[width=1.00\linewidth]{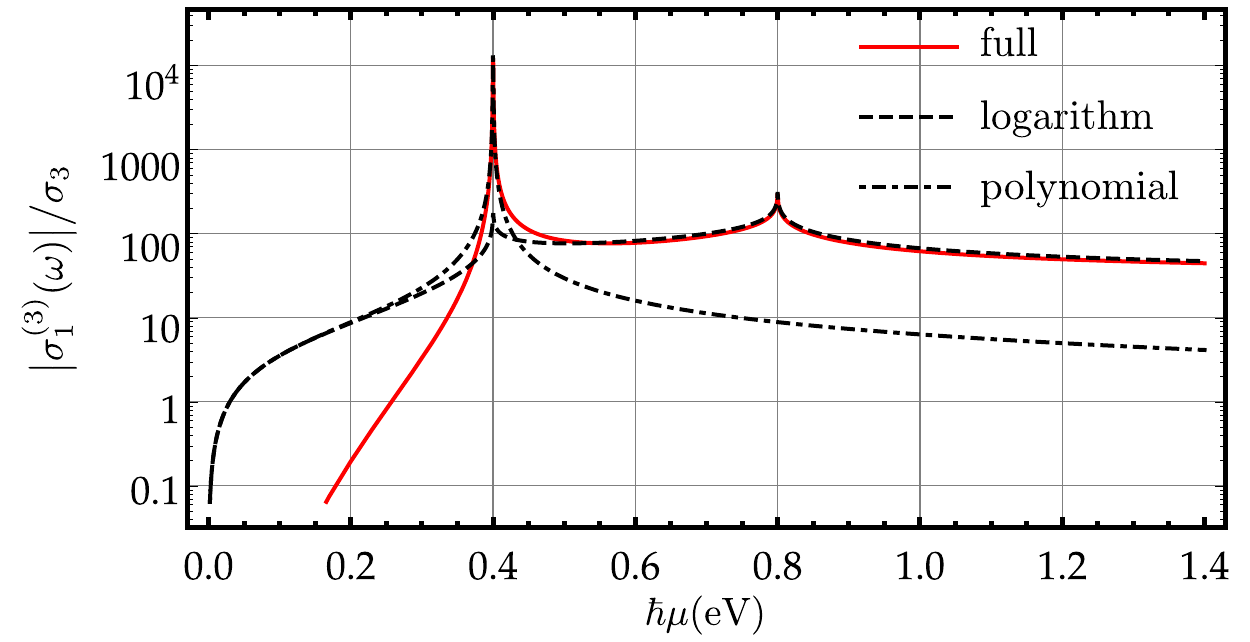}
\caption[Kerr effect in gapped graphene]{\label{fig:s1}%
Optical Kerr conductivity: the role of logarithmic divergences and polynomial 
resonances.
We set $\hbar\omega = 800$ meV in graphene ($\Delta\to 0$) as a function of 
doping.
Solid red curve represents the complete solution of \Eqref{eq:gapped:Kerr} and 
black curves represent the contributions from logarithmic divergences (dashed) 
and polynomial resonances (dot-dashed).}
\end{figure}

The polynomial resonances found in higher order contributions to a given 
Fourier component, \ie $j>n$, play a important role in the overall response of 
the system, by enhancing significantly the $(j+n-2)/2$ lowest resonances.
To illustrate this, we plot in \Fref{fig:s1} the contributions that arise 
from the logarithmic divergences and from the polynomial resonances in the 
optical Kerr conductivity $\sigma_1^{(3)}(\omega)$.
The effect is two-fold, first and most striking is the dramatic increase in the 
amplitude of the lowest resonance by virtue of the polynomial resonance.
Second is the destructive combination of both contributions for small doping
$\mu< \omega/2$.

%
%
%
%

\section{Higher order terms and role of doping}
\begin{figure}[b]
\includegraphics[width=1.00\linewidth]{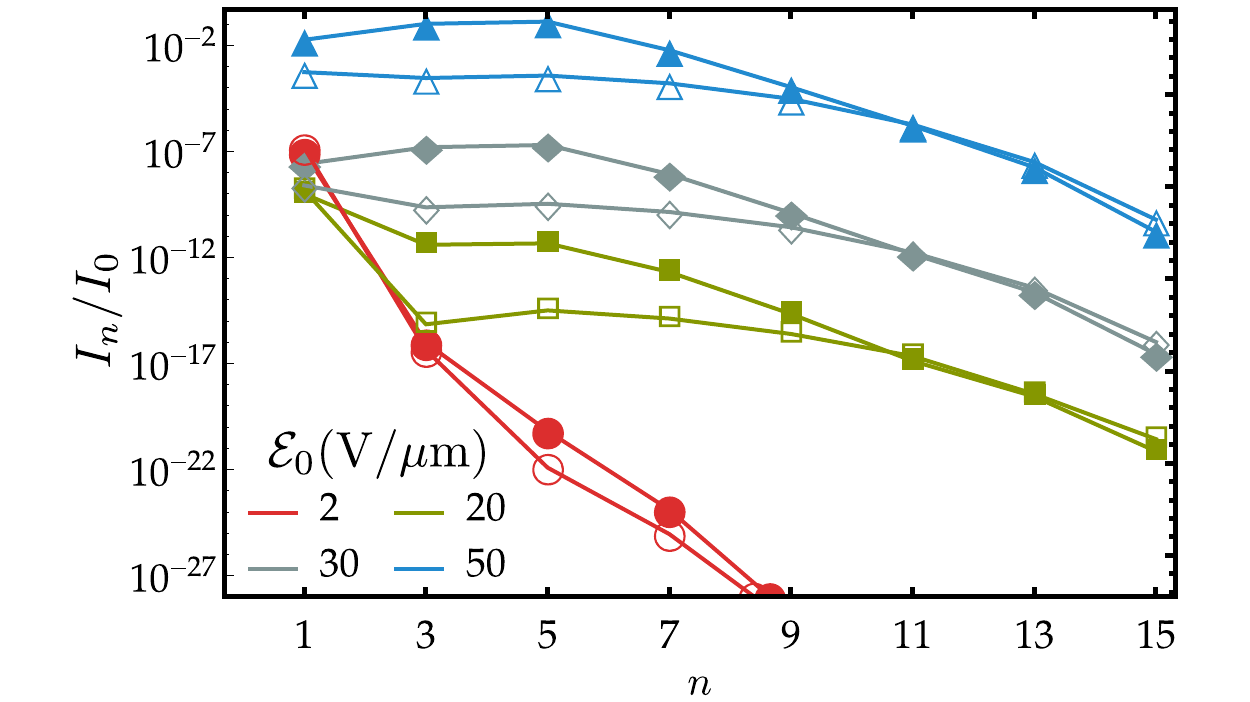}
\caption[Role of doping in nonlinear response]{\label{fig:s3}%
The role of doping in the nonlinear response.
Relative intensity of the $n^\mathrm{th}$ Fourier components of the field 
radiated by the current response in doped graphene, 
$\hbar\mu =  \{ 250, 312.5 \}$ meV, full and open symbols respectively, 
considering an external monochromatic field 
with photon energy $\hbar\omega = 100$ meV and intensity in vacuum
$I_0 = \varepsilon_0 c_0 \Ecal_0^2/2$.}
\end{figure}

Doping plays an important role in the optical response of graphene. At linear 
order its effect is well know, primarily blocking the real part of the response 
for $\omega < 2\mu$ and sets the amplitude of the Drude peak.
%
Yet, as discussed above and  on the main paper (see Fig. 2) its role in the 
nonlinear response can be drastic.
In \Fref{fig:s3}, we plot the amplitude of the Fourier components of the 
radiated field at two different doping levels 
$\hbar\mu =  \{ 250, 312.5 \}$ meV.
Despite the relatively small change in doping, the nonlinear response changes 
drastically, particularly the Fourier components $n=\{3,5,7\}$, stressing once 
more the importance of the interplay between doping level and photon energy.
%

%
%
%
%

\section{Conductivities up to ninth order for gapped systems}

The general expressions for the $n^\mathrm{th}$ harmonic and higher order 
corrections $j>n$ in gapped graphene can be expressed in the form
\begin{subequations}
\begin{align}
\label{eq:gapped:sigma:nn}
\sigma_n^{(n)}(\omega) &=
\frac{ i \sigma_n }{ \pi }
\bigg( \frac{ 2v_F }{ 3a_0 \omega } \bigg)^{2(n-1)}
%
\sum_{m=1}^{ n }  
\ln \frac{ 2\Omega -m \omega }{ 2\Omega +m \omega }
\nonumber\\ &
\times \sum_{k=0}^{ (n+1)/2 }
c_{km}^{(nn)} \bigg( \frac{ \Delta }{ \omega } \bigg)^{2k} ,
\end{align}
\begin{align}
\label{eq:gapped:sigma:nj}
\sigma_n^{(j)}(\omega) &=
\frac{ i \sigma_j }{ \pi }
\bigg( \frac{ 2v_F }{ 3a_0 \omega } \bigg)^{2(j-1)}
%
\bigg[ 
\sum_{m=1}^{ (n+j)/2 } 
\ln \frac{ 2\Omega -m \omega }{ 2\Omega +m \omega }
\nonumber\\ &
\times \sum_{k=0}^{ (j+1)/2 }
c_{km}^{(nj)} \bigg( \frac{ \Delta }{ \omega } \bigg)^{2k} 
+ \frac{ P_{nj} }{ Q_{nj} }
\bigg] \; (j > n) ,
\end{align}
\end{subequations}
where $c_{km}^{(nj)}$ are constant coefficients, $P_{nj}$ and 
$Q_{nj}$ are polynomial functions in $\omega$,
and recall the systematic definition of the nonlinear conductivity scales 
introduced in the main text 
$\sigma_{j>1} \equiv 2[3 e a_0 /(4 \hbar v_F)]^{j-1} \sigma_1$.
As shown in Eq. (11) in the main text, all nonlinear conductivities are regular 
at vanishing doping. The complete list of non-zero coefficients $q_{nj}$, up to 
ninth order, can be found in \Tbref{tab:qnj}.
%
%
%
\begin{table*}[t]
\caption{\label{tab:qnj}%
Coefficients $q_{nj}$ for nonlinear conductivities of graphene at the zero 
doping limit.}
\begin{ruledtabular}
\begin{tabular}{lccccc}
$n$ & $j=1$ & $3$       & $5$           & $7$   
    & $9$\\ \hline
$1$ & $1$   & $45/256$  & $1629/16384$  & $1708371/10485760$
    & $41182970949 /134217728000 $
\\
$3$ &       & $3/512$   & $-351/32768$  & $-30475581 /838860800$
    & $-11749230389433 /1342177280000 $
\\
$5$ &       &           & $-45/65536$   & $ 777122073/262144000 $
    & $ -1205821487213379/ 164416716800000 $
\\
$7$ &       &           &               & $7371/104857600$ 
    & $ -1251787974784857 /460366807040000 $
\\
$9$ &       &           &               &           
    & $-3971997/751619276800$ 
\end{tabular}
\end{ruledtabular}
\end{table*}
%
%

We provide the full expressions for the optical Kerr effect, fifth, seventh and 
ninth harmonic conductivities, namely \Eqsref{eq:gapped:Kerr}, 
\Pref{eq:gap:5HG}, \Pref{eq:gap:7HG} and \Pref{eq:gap:9HG}.
For $j>n$, we provide the coefficients $c_{km}^{(nj)}$ in \Tbref{tab:C1} and 
the respective polynomial functions $P_{nj}$ and $Q_{nj}$, found in 
\Eqsref{eq:P1J} and \Pref{eq:Q7J}.
Results for gapless graphene can be expressed directly from those of gapped 
graphene, by taking the limit $\Delta \to 0$ and replacing $\Omega$ by $\mu$.

The optical Kerr conductivity in gapped graphene reads
\begin{align}
\label{eq:gapped:Kerr}
\sigma_{1}^{(3)}( \omega) &= 
\frac{ 9i\sigma_3 }{ 256 \pi }
\bigg( \frac{ 2v_F }{ 3a_0 \omega } \bigg)^4
\bigg[
\nonumber\\&
\frac{ 4 [ 3 \Omega^2 \omega^4 +8 \Delta^2 \Omega^2 \omega^2
+48\Delta^4( 9\Omega^2 -2\omega^2 ) ]
}{ \omega^3 \mu ( 4\mu^2 -\omega^2 )}
\nonumber\\&
+\bigg( 16 +\frac{32\Delta^2}{\omega^2} -\frac{48\Delta^4}{\omega^4} \bigg)
\ln \frac{2\omega-2\Omega}{2\omega+2\Omega}
\nonumber\\&
-\bigg( 11 +\frac{56\Delta^2}{\omega^2} -\frac{5283\Delta^4}{\omega^4} \bigg)
\ln \frac{\omega-2\Omega}{\omega+2\Omega}
\bigg] .
\end{align}

\begin{widetext}
The fifth, seventh and ninth harmonic conductivities read
{ \footnotesize
\begin{align}
\label{eq:gap:5HG}
\sigma_{5}^{(5)} (\omega) &= 
\frac{ 9i \sigma_5 }{ 655360 \pi } 
\bigg( \frac{ 2v_F }{ 3a_0 \omega } \bigg)^8
\bigg[
   \bigg( 4925
    +\frac{ 3788 \Delta^2 }{ \omega^2 } 
    -\frac{ 2640 \Delta^4 }{ \omega^4 }
    +\frac{  320 \Delta^6 }{ \omega^6 }
  \bigg)
  \ln \frac{2\Omega-5\omega}{2\Omega+5\omega}
\nonumber\\&
  -\bigg(   16384
    +\frac{ 16384 \Delta^2 }{ \omega^2 }
    -\frac{ 15360 \Delta^4 }{ \omega^4 }
    +\frac{  2560 \Delta^6 }{ \omega^6 }
  \bigg)
  \ln \frac{2\Omega-4\omega}{2\Omega+4\omega}
%
%
  +\bigg(   19359
    +\frac{ 27396 \Delta^2 }{ \omega^2 } 
    -\frac{ 36720 \Delta^4 }{ \omega^4 }
    +\frac{  8640 \Delta^6 }{ \omega^6 }
  \bigg)
  \ln \frac{2\Omega-3\omega}{2\Omega+3\omega}
\nonumber\\&
  -\bigg(    9216
    +\frac{ 21504 \Delta^2 }{ \omega^2 }
    -\frac{ 46080 \Delta^4 }{ \omega^4 }
    +\frac{ 15360 \Delta^6 }{ \omega^6 }
  \bigg)
  \ln \frac{2\Omega-2\omega}{2\Omega+2\omega}
%
%
  +\bigg(    1266
    +\frac{  7416 \Delta^2 }{ \omega^2 }
    -\frac{ 30240 \Delta^4 }{ \omega^4 }
    +\frac{ 13440 \Delta^6 }{ \omega^6 } 
  \bigg)
   \ln \frac{2\Omega- \omega}{2\Omega+ \omega}
\bigg] ,
\end{align}
%
%
\begin{align}
\label{eq:gap:7HG}
\sigma_{7}^{(7)} (\omega) &=  
\frac{ i \sigma_7 }{ 2936012800 \pi } 
\bigg( \frac{ 2v_F }{ 3a_0 \omega } \bigg)^{12}
\bigg[
%
%
  -\bigg(   5126429
    +\frac{ 2911408\Delta^2 }{ \omega^2 } 
    -\frac{ 1680672\Delta^4 }{ \omega^4 }
    +\frac{  224000\Delta^6 }{ \omega^6 }
    -\frac{    8960\Delta^8 }{ \omega^8 }
  \bigg)
  \ln \frac{2\Omega-7\omega}{2\Omega+7\omega}
\nonumber\\&
  +\bigg(   27122688
    +\frac{ 18468864 \Delta^2 }{ \omega^2 } 
    -\frac{ 13031424 \Delta^4 }{ \omega^4 }
    +\frac{  2150400 \Delta^6 }{ \omega^6 }
    -\frac{   107520 \Delta^8 }{ \omega^8 }
  \bigg)
  \ln \frac{2\Omega-6\omega}{2\Omega+6\omega}
\nonumber\\&
  -\bigg(   58448125
    +\frac{ 49425200\Delta^2 }{ \omega^2 } 
    -\frac{ 43925280\Delta^4 }{ \omega^4 }
    +\frac{  9184000\Delta^6 }{ \omega^6 }
    -\frac{   582400\Delta^8 }{ \omega^8 }
  \bigg)
  \ln \frac{2\Omega-5\omega}{2\Omega+5\omega}
\nonumber\\&
  +\bigg(   65011712
    +\frac{ 71827456 \Delta^2 }{ \omega^2 } 
    -\frac{ 83951616 \Delta^4 }{ \omega^4 }
    +\frac{ 22937600 \Delta^6 }{ \omega^6 }
    -\frac{ 1863680\Delta^8 }{ \omega^8 }
  \bigg)
  \ln \frac{2\Omega-4\omega}{2\Omega+4\omega}
\nonumber\\&
  -\bigg(   38624121
    +\frac{ 60454512 \Delta^2 }{ \omega^2 } 
    -\frac{ 99273888 \Delta^4 }{ \omega^4 }
    +\frac{ 36960000 \Delta^6 }{ \omega^6 }
    -\frac{  3843840 \Delta^8 }{ \omega^8 }
  \bigg)
  \ln \frac{2\Omega-3\omega}{2\Omega+3\omega}
\nonumber\\&
  +\bigg(   11127808
    +\frac{ 28676096 \Delta^2 }{ \omega^2 } 
    -\frac{ 74102784 \Delta^4 }{ \omega^4 }
    +\frac{ 39424000 \Delta^6 }{ \omega^6 }
    -\frac{ 5125120  \Delta^8 }{ \omega^8 }
  \bigg)
  \ln \frac{2\Omega-2\omega}{2\Omega+2\omega}
\nonumber\\&
  -\bigg(   1040601
    +\frac{  6605808 \Delta^2 }{ \omega^2 } 
    -\frac{ 32987808 \Delta^4 }{ \omega^4 }
    +\frac{ 25132800 \Delta^6 }{ \omega^6 }
    -\frac{  3843840 \Delta^8 }{ \omega^8 }
  \bigg)
   \ln \frac{2\Omega- \omega}{2\Omega+ \omega}
\bigg] ,
\end{align}
%
%
\begin{align}
\label{eq:gap:9HG}
\sigma_{9}^{(9)} (\omega) &=  
\frac{ 27 i \sigma_9 }{ 10522669875200 \pi } 
\bigg( \frac{ 2v_F }{ 3a_0 \omega } \bigg)^{16}
\bigg[
\nonumber\\&
   \bigg(   811948293
    +\frac{ 369488004 \Delta^2    }{ \omega^2    } 
    -\frac{ 182320480 \Delta^4    }{ \omega^4    }
    +\frac{  23327360 \Delta^6    }{ \omega^6    }
%
%
    -\frac{   1191680 \Delta^8    }{ \omega^8    }
    +\frac{     21504 \Delta^{10} }{ \omega^{10} }
  \bigg)
  \ln \frac{2\Omega-9\omega}{2\Omega+9\omega}
\nonumber\\&
  -\bigg(   5905580032
    +\frac{ 3087007744 \Delta^2    }{ \omega^2    } 
    -\frac{ 1772093440 \Delta^4    }{ \omega^4    }
    +\frac{  266076160 \Delta^6    }{ \omega^6    }
    -\frac{   16056320 \Delta^8    }{ \omega^8    }
%
%
    +\frac{     344064 \Delta^{10} }{ \omega^{10} }
  \bigg)
  \ln \frac{2\Omega-8\omega}{2\Omega+8\omega}
\nonumber\\&
  +\bigg(   18559737203
    +\frac{ 11360173916 \Delta^2    }{ \omega^2    } 
    -\frac{  7720364960 \Delta^4    }{ \omega^4    }
    +\frac{  1381063040 \Delta^6    }{ \omega^6    }
    -\frac{    99662080 \Delta^8    }{ \omega^8    }
%
%
    +\frac{     2558976 \Delta^{10} }{ \omega^{10} }
  \bigg)
  \ln \frac{2\Omega-7\omega}{2\Omega+7\omega}
\nonumber\\&
  -\bigg(   32794509312
    +\frac{ 24102076416 \Delta^2    }{ \omega^2    } 
    -\frac{ 19837419520 \Delta^4    }{ \omega^4    }
    +\frac{  4309975040 \Delta^6    }{ \omega^6    }
    -\frac{   377323520 \Delta^8    }{ \omega^8    }
%
%
    +\frac{    11698176 \Delta^{10} }{ \omega^{10} }
  \bigg)
  \ln \frac{2\Omega-6\omega}{2\Omega+6\omega}
\nonumber\\&
  +\bigg(   35426172500
    +\frac{ 32354387600 \Delta^2    }{ \omega^2    } 
    -\frac{ 33255536000 \Delta^4    }{ \omega^4    }
    +\frac{  8998976000 \Delta^6    }{ \omega^6    }
    -\frac{   972160000 \Delta^8    }{ \omega^8    }
%
%
    +\frac{    36556800 \Delta^{10} }{ \omega^{10} }
  \bigg)
  \ln \frac{2\Omega-5\omega}{2\Omega+5\omega}
\nonumber\\&
  -\bigg(   23722983424
    +\frac{ 28303163392 \Delta^2    }{ \omega^2    } 
    -\frac{ 37947965440 \Delta^4    }{ \omega^4    }
    +\frac{ 13230407680 \Delta^6    }{ \omega^6    }
    -\frac{  1798307840 \Delta^8    }{ \omega^8    }
%
%
    +\frac{    81887232 \Delta^{10} }{ \omega^{10} }
  \bigg)
  \ln \frac{2\Omega-4\omega}{2\Omega+4\omega}
\nonumber\\&
  +\bigg(   9439968636
    +\frac{ 15933700272 \Delta^2    }{ \omega^2    }
    -\frac{ 29786529920 \Delta^4    }{ \omega^4    }
    +\frac{ 14022561280 \Delta^6    }{ \omega^6    }
    -\frac{ 2442818560  \Delta^8    }{ \omega^8    }
%
%
    +\frac{ 133066752   \Delta^{10} }{ \omega^{10} } 
  \bigg)
  \ln \frac{2\Omega-3\omega}{2\Omega+3\omega}
\nonumber\\&
  -\bigg(   1956020224
    +\frac{  5417893888 \Delta^2    }{ \omega^2    } 
    -\frac{ 15830548480 \Delta^4    }{ \omega^4    }
    +\frac{ 10704977920 \Delta^6    }{ \omega^6    }
    -\frac{  2400419840 \Delta^8    }{ \omega^8    }
%
%
    +\frac{   152076288 \Delta^{10} }{ \omega^{10} }
  \bigg)
  \ln \frac{2\Omega-2\omega}{2\Omega+2\omega}
\nonumber\\&
  +\bigg(   139206806
    +\frac{  937313528 \Delta^2    }{ \omega^2    } 
    -\frac{ 5333514560 \Delta^4    }{ \omega^4    }
    +\frac{ 5380094720 \Delta^6    }{ \omega^6    }
    -\frac{ 1488847360 \Delta^8    }{ \omega^8    }
%
%
    +\frac{  104552448 \Delta^{10} }{ \omega^{10} }
  \bigg)
   \ln \frac{2\Omega- \omega}{2\Omega+ \omega}
\bigg] .
\end{align}
}
The polynomial functions of the family $P_{nj}$ read
{ \footnotesize
\begin{subequations}
\label{eq:P1J}
\begin{align}
P_{15} &= 
-27 \Big[
    -320 \Delta ^6 \Big(
         6860 \mu ^8
        -9795 \mu ^6 \omega ^2
        +3149 \mu ^4 \omega ^4
         -216 \mu ^2 \omega ^6
          -16 \omega ^8 \Big)
    +48 \Delta ^4 \Big(
         4340 \mu ^8 \omega ^2
        -6469 \mu ^6 \omega ^4
        +2297 \mu ^4 \omega ^6
         -240 \mu ^2 \omega ^8 \Big)
\nonumber\\ &
    -4 \Delta ^2 \Big(
           4 \mu ^8 \omega ^4
        +343 \mu ^6 \omega ^6
         -59 \mu ^4 \omega ^8 \Big)
    -1276 \mu ^8 \omega ^6
     +119 \mu ^6 \omega ^8
       +5 \mu ^4 \omega ^{10} \Big] ,
\\
P_{17} &= 
81 \Big[
    896 \Delta ^8 \Big(
         220725120 \mu ^{16}
        -1085717600 \mu ^{14} \omega^2
        +1871794584 \mu ^{12} \omega^4
        -1439294370 \mu ^{10} \omega^6
        +506082893 \mu ^8 \omega ^8
        -77041635 \mu ^6 \omega ^{10}
\nonumber\\ &
        +3399312 \mu ^4 \omega ^{12}
        +20160 \mu ^2 \omega ^{14}
        +20736 \omega ^{16}\Big)
    -1280 \Delta ^6 \mu ^2 \omega ^2 \Big(
          24602816 \mu ^{14}
        -121233232 \mu ^{12} \omega ^2
        +209561348 \mu ^{10} \omega ^4
        -161682431 \mu ^8 \omega ^6
\nonumber\\ &
         +57044056 \mu ^6 \omega ^8
          -8625293 \mu ^4 \omega ^{10}
           +300768 \mu ^2 \omega ^{12}
            +27648 \omega ^{14} \Big)
    +96 \Delta ^4 \mu ^4 \omega ^4 \Big(
          8962240 \mu ^{12}
        -44433104 \mu ^{10} \omega ^2
        +77619556 \mu ^8 \omega ^4
\nonumber\\ &
        -60976999 \mu ^6 \omega ^6
        +22423706 \mu ^4 \omega ^8
         -3822343 \mu ^2 \omega ^{10}
          +244224 \omega ^{12} \Big)
    -16 \Delta ^2 \mu ^6 \omega ^6 \Big(
        321472 \mu ^{10}
        -1125392 \mu ^8 \omega ^2
        +964020 \mu ^6 \omega ^4
\nonumber\\ &
        -259267 \mu ^4 \omega ^6
        +27338 \mu ^2 \omega ^8
        +2709 \omega ^{10}\Big)
    +\mu ^6 \omega ^8 \Big(
        -4269376 \mu ^{10}
        +16069552 \mu ^8 \omega^2
        -16274748 \mu ^6 \omega^4
        +7469521 \mu ^4 \omega ^6
\nonumber\\ &
        -1748206 \mu ^2 \omega^8
        +135657 \omega ^{10}\Big)
    \Big] ,
\\
P_{19} &= 
27 \Big[
    46080 \Delta ^{10} \Big(187256945075200 \mu ^{26}
        -2325179092531456  \mu ^{24} \omega ^2
        +11687240067038592 \mu ^{22} \omega ^4
        -31227635973633056 \mu ^{20} \omega ^6
\nonumber \\ &
        +48860300608451540 \mu ^{18} \omega ^8
        -46478022674034021 \mu ^{16} \omega ^{10}
        +26989978924387315 \mu ^{14} \omega ^{12}
        -9364370227141651  \mu ^{12} \omega ^{14}
\nonumber \\ &
        +1846375279461405  \mu ^{10} \omega ^{16}
        -181530914846796   \mu ^8 \omega ^{18}
        +5430654668160     \mu ^6 \omega ^{20}
        +169113692160      \mu ^4 \omega ^{22}
        -15282266112       \mu ^2 \omega ^{24}
\nonumber \\ &
        +3224862720 \omega ^{26}\Big)
    -8960 \Delta ^8 \mu ^2 \omega ^2 \Big(
        207124212399104     \mu ^{24}
        -2572102733782784   \mu ^{22} \omega ^2
        +12929922371094144  \mu ^{20} \omega ^4
\nonumber \\ &
        -34553358155256544  \mu ^{18} \omega ^6
        +54074328703809484  \mu ^{16} \omega ^8
        -51449663244720339  \mu ^{14} \omega ^{10}
        +29884780216342349  \mu ^{12} \omega ^{12}
\nonumber \\ &
        -10371651232074029  \mu ^{10} \omega ^{14}
        +2046000204112899   \mu ^8 \omega ^{16}
        -202081754607804    \mu ^6 \omega ^{18}
        +6770075385600      \mu ^4 \omega ^{20}
\nonumber \\ &
        -102269952000       \mu ^2 \omega ^{22}
        +35473489920 \omega ^{24} \Big)
    +3200 \Delta ^6 \mu ^4 \omega ^4 \Big(
        31610284073984      \mu ^{22}
        -392635665455360    \mu ^{20} \omega ^2
\nonumber \\ &
        +1974400862745984 \mu ^{18} \omega ^4
        -5278454843537440 \mu ^{16} \omega ^6
        +8264714267791444 \mu ^{14} \omega ^8
        -7868286992375685 \mu ^{12} \omega ^{10}
\nonumber \\ &
        +4573179204764219 \mu ^{10} \omega ^{12}
        -1587286528597115 \mu ^8 \omega ^{14}
        +311808898399989 \mu ^6 \omega ^{16}
        -29677946203620 \mu ^4 \omega ^{18}
\nonumber \\ &
        +559840481280 \mu ^2 \omega ^{20}
        +67124920320 \omega ^{22}\Big)
    -96 \Delta ^4 \mu ^6 \omega ^6 \Big(
        12461343792128 \mu ^{20}
        -153934549591808 \mu ^{18} \omega ^2
        +768680423211648 \mu ^{16} \omega ^4
\nonumber \\ &
        -2037737318722528 \mu ^{14} \omega ^6
        +3159909982948588 \mu ^{12} \omega ^8
        -2978876192190843 \mu ^{10} \omega ^{10}
        +1718919063655493 \mu ^8 \omega ^{12}
\nonumber \\ &
        -598137631786373 \mu ^6 \omega ^{14}
        +121291455245643 \mu ^4 \omega ^{16}
        -13131969585948 \mu ^2 \omega ^{18}
        +585252864000 \omega ^{20}\Big)
\nonumber \\ &
    +4 \Delta ^2 \mu ^8 \omega ^8 \Big(
        11128861893632 \mu ^{18}
        -128298690021632 \mu ^{16} \omega ^2
        +584308470525312 \mu ^{14} \omega ^4
        -1375544099233312 \mu ^{12} \omega ^6
\nonumber \\ &
        +1838156397364372 \mu ^{10} \omega ^8
        -1446756964610997 \mu ^8 \omega ^{10}
        +672490048294667 \mu ^6 \omega ^{12}
        -180284596846667 \mu ^4 \omega ^{14}
\nonumber \\ &
        +25742523524517 \mu ^2 \omega ^{16}
        -1539147689892 \omega ^{18}\Big)
    +\mu ^8 \omega ^{10} \Big(
        16473311681536 \mu ^{18}
        -185685695854336 \mu ^{16} \omega ^2
        +820185567144576 \mu ^{14} \omega ^4
\nonumber \\ &
        -1854452838481376 \mu ^{12} \omega ^6
        +2358902774973356 \mu ^{10} \omega ^8
        -1755059863889931 \mu ^8 \omega ^{10}
        +760761649666741 \mu ^6 \omega ^{12}
        -187368464362741 \mu ^4 \omega ^{14}
\nonumber \\ &
        +24210354003291 \mu ^2 \omega ^{16}
        -1311096961116 \omega ^{18}
    \Big) 
\Big] ,
\\
P_{35} &= 
9 \Big[
    1600 \Delta ^6 \Big(
        140     \mu ^6
        -495    \mu ^4 \omega ^2
        +439    \mu ^2 \omega ^4
        -72 \omega ^6\Big)
    +48 \Delta ^4 \omega ^2 \Big(
        -1060   \mu ^6
        +3837   \mu ^4 \omega ^2
        -3593   \mu ^2 \omega ^4
        +576    \omega ^6\Big)
\nonumber \\ &
    +4 \Delta ^2 \Big(
        436     \mu ^6 \omega ^4
        -1009   \mu ^4 \omega ^6
        -387    \mu ^2 \omega ^8\Big)
    +\mu ^2 \omega ^6 \Big(
        2572    \mu ^4
        -7231   \mu ^2 \omega ^2
        +819    \omega ^4
    \Big)
\Big] ,
\\
P_{37} &= 
9 \Big[
    8960 \Delta ^8 \Big(
        154842688 \mu ^{16}
        -1693862192 \mu ^{14} \omega ^2
        +7076655628 \mu ^{12} \omega ^4
        -14432352001 \mu ^{10} \omega ^6
        +15117336290 \mu ^8 \omega ^8
        -7801203157 \mu ^6 \omega ^{10}
\nonumber \\ &
        +1676409144 \mu ^4 \omega ^{12}
        -92253600 \mu ^2 \omega ^{14}
        -6220800 \omega ^{16}\Big)
    +6400 \Delta ^6 \omega ^2 \Big(
        -78564416 \mu ^{16}
        +861403824 \mu ^{14} \omega ^2
        -3610618540 \mu ^{12} \omega ^4
\nonumber \\ &
        +7397960105 \mu ^{10} \omega ^6
        -7800557442 \mu ^8 \omega ^8
        +4065811129 \mu ^6 \omega ^{10}
        -894421668 \mu ^4 \omega ^{12}
        +57514752 \mu ^2 \omega ^{14}
        +1990656 \omega ^{16}\Big)
\nonumber \\ &
    +96 \Delta ^4 \mu ^2 \omega ^4 \Big(
        240655296 \mu ^{14}
        -2644591184 \mu ^{12} \omega ^2
        +11132594516 \mu ^{10} \omega ^4
        -22987069287 \mu ^8 \omega ^6
        +24527909390 \mu ^6 \omega ^8
        -13001424359 \mu ^4 \omega ^{10}
\nonumber \\ &
        +2981172348 \mu ^2 \omega ^{12}
        -238878720 \omega ^{14}\Big)
    -16 \Delta ^2 \mu ^4 \omega ^6 \Big(
        88161088    \mu ^{12}
        -833246704  \mu ^{10} \omega ^2
        +2795912700 \mu ^8 \omega ^4
        -4054626277 \mu ^6 \omega ^6
\nonumber \\ &
        +2413067146 \mu ^4 \omega ^8
        -488748069  \mu ^2 \omega ^{10}
        +38008116   \omega ^{12}\Big)
    +\mu ^4 \omega ^8 \Big(
        -1223424832  \mu ^{12}
        +11825233648 \mu ^{10} \omega ^2
        -41147030652 \mu ^8 \omega ^4
\nonumber \\ &
        +62929802869 \mu ^6 \omega ^6
        -41214902890 \mu ^4 \omega ^8
        +10669588533 \mu ^2 \omega ^{10}
        -1009826676  \omega ^{12}
    \Big)
\Big] ,
\\
P_{39} &= 
9 \Big[
    64512 \Delta ^{10} \Big(
        309753515495424 \mu ^{28}
        -7641740299675648 \mu ^{26} \omega ^2
        +80412581120246528 \mu ^{24} \omega ^4
        -475750643401113088 \mu ^{22} \omega ^6
\nonumber \\ &
        +1754584695596524208 \mu ^{20} \omega ^8
        -4223937570253353928 \mu ^{18} \omega ^{10}
        +6751925001676414493 \mu ^{16} \omega ^{12}
        -7145124722011657483 \mu ^{14} \omega ^{14}
\nonumber \\ &
        +4898503653386626283 \mu ^{12} \omega ^{16}
        -2080663453422856033 \mu ^{10} \omega ^{18}
        +505895262262389564 \mu ^8 \omega ^{20}
        -60849869183824320 \mu ^6 \omega ^{22}
\nonumber \\ &
        +2320230220416000 \mu ^4 \omega ^{24}
        +4237671168000 \mu ^2 \omega ^{26}
        +12093235200000 \omega ^{28}\Big)
    -8960 \Delta ^8 \omega ^2 \Big(
        972060393148416 \mu ^{28}
\nonumber \\ &
        -23984147489032192 \mu ^{26} \omega ^2
        +252421208707688192 \mu ^{24} \omega ^4
        -1493720153089864192 \mu ^{22} \omega ^6
        +5510291747508277232 \mu ^{20} \omega ^8
\nonumber \\ &
        -13269530175856290472 \mu ^{18} \omega ^{10}
        +21219569657598491417 \mu ^{16} \omega ^{12}
        -22466131756544682187 \mu ^{14} \omega ^{14}
        +15411021721264809947 \mu ^{12} \omega ^{16}
\nonumber \\ &
        -6549877232232901057 \mu ^{10} \omega ^{18}
        +1592687514214980096 \mu ^8 \omega ^{20}
        -190486298653441200 \mu ^6 \omega ^{22}
        +6526133560627200 \mu ^4 \omega ^{24}
\nonumber \\ &
        +218411889868800 \mu ^2 \omega ^{26}
        +19349176320000 \omega ^{28}\Big)
    +640 \Delta ^6 \mu ^2 \omega ^4 \Big(
        1222308154912768 \mu ^{26}
        -30167798532724736 \mu ^{24} \omega ^2
\nonumber \\ &
        +317620928184714496 \mu ^{22} \omega ^4
        -1880414993390921216 \mu ^{20} \omega ^6
        +6940699972336059856 \mu ^{18} \omega ^8
        -16725442627223072696 \mu ^{16} \omega ^{10}
\nonumber \\ &
        +26767586681448735451 \mu ^{14} \omega ^{12}
        -28367813457999362681 \mu ^{12} \omega ^{14}
        +19484307833540761081 \mu ^{10} \omega ^{16}
        -8298098342699425131 \mu ^8 \omega ^{18}
\nonumber \\ &
        +2026708139492503848 \mu ^6 \omega ^{20}
        -245089631574581040 \mu ^4 \omega ^{22}
        +8437345689600000 \mu ^2 \omega ^{24}
        +451480780800000 \omega ^{26}\Big)
\nonumber \\ &
    -480 \Delta ^4 \mu ^4 \omega ^6 \Big(
        33211897040896 \mu ^{24}
        -812271998609408 \mu ^{22} \omega ^2
        +8458992945701632 \mu ^{20} \omega^4
        -49438723339383296 \mu ^{18} \omega ^6
\nonumber \\ &
        +179773666199270320 \mu ^{16} \omega ^8
        -425870027861226824 \mu ^{14} \omega ^{10}
        +668539751184633373 \mu ^{12} \omega ^{12}
        -693525260286524207 \mu ^{10} \omega ^{14}
\nonumber \\ &
        +465737013219803311 \mu ^8 \omega ^{16}
        -194382587913945165 \mu ^6 \omega ^{18}
        +47152930812034968 \mu ^4 \omega ^{20}
        -5973952612395600 \mu ^2 \omega ^{22}
\nonumber \\ &
        +300987187200000 \omega ^{24}\Big)
    +4 \Delta ^2 \mu ^6 \omega ^8 \Big(
        536217522040832 \mu ^{22}
        -12496633825982464 \mu ^{20} \omega ^2
        +122387461446996224 \mu ^{18} \omega ^4
\nonumber \\ &
        -661840153597033984 \mu ^{16} \omega ^6
        +2181946355016020624 \mu ^{14} \omega ^8
        -4567485492086789464 \mu ^{12} \omega ^{10}
        +6132062217877668239 \mu ^{10} \omega ^{12}
\nonumber \\ &
        -5217351621756729589 \mu ^8 \omega ^{14}
        +2723231065199668469 \mu ^6 \omega ^{16}
        -822367121180168799 \mu ^4 \omega ^{18}
        +130218923351183112 \mu ^2 \omega ^{20}
\nonumber \\ &
        -8214161326873200 \omega ^{22}\Big)
    +\mu ^6 \omega ^{10} \Big(
        1326111826087936 \mu ^{22}
        -30780984984061952 \mu ^{20} \omega ^2
        +299899605136498432 \mu ^{18} \omega ^4
\nonumber \\ &
        -1611098752492186112 \mu ^{16} \omega ^6
        +5266492798707198832 \mu ^{14} \omega ^8
        -10904888055843848552 \mu ^{12} \omega ^{10}
        +14444414998165145377 \mu ^{10} \omega ^{12}
\nonumber \\ &
        -12103980506737587227 \mu ^8 \omega ^{14}
        +6224529344696909467 \mu ^6 \omega ^{16}
        -1854359645704647057 \mu ^4 \omega ^{18}
        +290123422224942456 \mu ^2 \omega ^{20}
\nonumber \\ &
        -18166817810451600 \omega ^{22}
    \Big)
\Big] ,
\\
P_{57} &= 
-81 \Big[
    -62720 \Delta ^8 \Big(
        1232 \mu ^{10}
        -17016 \mu ^8 \omega ^2
        +79737 \mu ^6 \omega ^4
        -150974 \mu ^4 \omega ^6
        +105921 \mu ^2 \omega ^8
        -16200 \omega ^{10}\Big)
\nonumber \\ &
    +44800 \Delta ^6 \omega ^2 \Big(
        3184 \mu ^{10}
        -44088 \mu ^8 \omega ^2
        +207555 \mu ^6 \omega ^4
        -396607 \mu ^4 \omega ^6
        +283956 \mu ^2 \omega ^8
        -43200 \omega ^{10}\Big)
\nonumber \\ &
    +96 \Delta ^4 \omega ^4 \Big(
        -163408 \mu ^{10}
        +2304744 \mu ^8 \omega ^2
        -11210793 \mu ^6 \omega ^4
        +22779541 \mu ^4 \omega ^6
        -18591684 \mu ^2 \omega ^8
\nonumber \\ &
        +2764800 \omega ^{10}\Big)
    +16 \Delta ^2 \mu ^2 \omega ^6 \Big(
        103184 \mu ^8
        -1342344 \mu ^6 \omega ^2
        +5578101 \mu ^4 \omega ^4
        -7907441 \mu ^2 \omega ^6
        -60300 \omega ^8\Big)
\nonumber \\ &
    +\mu ^2 \omega ^8 \Big(
        1411888 \mu ^8
        -18802008 \mu ^6 \omega ^2
        +82255407 \mu ^4 \omega ^4
        -139348387 \mu ^2 \omega ^6
        +21260700 \omega ^8
    \Big)
\Big] ,
\\
P_{59} &= 
81 \Big[
    193536 \Delta ^{10} \Big(
        95335871488 \mu ^{24}
        -3474467149056 \mu ^{22} \omega ^2
        +53617091508608 \mu ^{20} \omega ^4
        -459325309409696 \mu ^{18} \omega ^6
\nonumber \\ &
        +2406126551013348 \mu ^{16} \omega ^8
        -8003718707211601 \mu ^{14} \omega ^{10}
        +17005840241750653 \mu ^{12} \omega ^{12}
        -22640508616620861 \mu ^{10} \omega ^{14}
\nonumber \\ &
        +18005536929127403 \mu ^8 \omega ^{16}
        -7798933351991286 \mu ^6 \omega ^{18}
        +1518142836231000 \mu ^4 \omega ^{20}
        -78654784320000 \mu ^2 \omega ^{22} 
        -5143824000000 \omega ^{24}\Big)
\nonumber \\ &
    +62720 \Delta ^8 \omega ^2 \Big(
        -598343263232 \mu ^{24}
        +21811826077952 \mu ^{22} \omega ^2
        -336705405969792 \mu ^{20} \omega ^4
        +2885724235393312 \mu ^{18} \omega ^6
\nonumber \\ &
        -15125197416086932 \mu ^{16} \omega ^8
        +50350676794142877 \mu ^{14} \omega ^{10}
        -107091761127166907 \mu ^{12} \omega ^{12}
        +142772361431828627 \mu ^{10} \omega ^{14}
\nonumber \\ &
        -113757742393642437 \mu ^8 \omega ^{16}
        +49407479442609732 \mu ^6 \omega ^{18}
        -9674380682323200 \mu ^4 \omega ^{20}
        +522367574400000 \mu ^2 \omega ^{22}
\nonumber \\ &
        +29393280000000 \omega ^{24}\Big)
    -4480 \Delta ^6 \omega ^4 \Big(
        -2021660253184 \mu ^{24}
        +73794708322048 \mu ^{22} \omega ^2
        -1141127989167744 \mu ^{20} \omega ^4
\nonumber \\ &
        +9802241047129568 \mu ^{18} \omega ^6
        -51531857164633964 \mu ^{16} \omega ^8
        +172232438007379083 \mu ^{14} \omega ^{10}
        -368281343263186729 \mu ^{12} \omega ^{12}
\nonumber \\ &
        +494471670546894313 \mu ^{10} \omega ^{14}
        -397667689343865879 \mu ^8 \omega ^{16}
        +174906664504972488 \mu ^6 \omega ^{18}
        -35123521332150000 \mu ^4 \omega ^{20}
\nonumber \\ &
        +2226207836160000 \mu ^2 \omega ^{22}
        +53747712000000 \omega ^{24}\Big)
    -96 \Delta ^4 \mu ^2 \omega ^6 \Big(
        3797717957632 \mu ^{22}
        -137353886787328 \mu ^{20} \omega ^2
\nonumber \\ &
        +2100673426799232 \mu ^{18} \omega ^4
        -17812683917478368 \mu ^{16} \omega ^6
        +92275903921064012 \mu ^{14} \omega ^8
        -303537823856362203 \mu ^{12} \omega ^{10}
\nonumber \\ &
        +638808180919849177 \mu ^{10} \omega ^{12}
        -845617510364262553 \mu ^8 \omega ^{14}
        +672165401963704647 \mu ^6 \omega ^{16}
        -292283205453447048 \mu ^4 \omega ^{18}
\nonumber \\ &
        +58403636668162800 \mu ^2 \omega ^{20}
        -4138573824000000 \omega ^{22}\Big)
    +4 \Delta ^2 \mu ^4 \omega ^8 \Big(
        22851662222336 \mu ^{20}
        -794742870529280 \mu ^{18} \omega ^2
\nonumber \\ &
        +11536344229520256 \mu ^{16} \omega ^4
        -91165811890041760 \mu ^{14} \omega ^6
        +428772403776990916 \mu ^{12} \omega ^8
        -1232248577574701265 \mu ^{10} \omega ^{10}
\nonumber \\ &
        +2137943903760337931 \mu ^8 \omega ^{12}
        -2133108076863341195 \mu ^6 \omega ^{14}
        +1110660721547411061 \mu ^4 \omega ^{16}
        -256938447392559000 \mu ^2 \omega ^{18}
\nonumber \\ &
        +21939596325090000 \omega ^{20}\Big)
    +\mu ^4 \omega ^{10} \Big(
        60324192345088 \mu ^{20}
        -2104270345527040 \mu ^{18} \omega ^2
        +30665663343838848 \mu ^{16} \omega ^4
\nonumber \\ &
        -243615590722170080 \mu ^{14} \omega ^6
        +1154223486667714028 \mu ^{12} \omega ^8
        -3351601703029398195 \mu ^{10} \omega ^{10}
        -5989529637878143585 \mu ^6 \omega ^{14}
\nonumber \\ &
        +5896994399270064673 \mu ^8 \omega ^{12}
        +3195062570596924863 \mu ^4 \omega ^{16}
        -775441141395078600 \mu ^2 \omega ^{18}
        +69308496112230000 \omega ^{20}
    \Big)
\Big] ,
\\
P_{79} &= 
243 \Big[
    9225216 \Delta ^{10} \Big(
        8640 \mu ^{14}
        -303056 \mu ^{12} \omega ^2
        +4051828 \mu ^{10} \omega ^4
        -26153487 \mu ^8 \omega ^6
        +85235078 \mu ^6 \omega ^8
        -133654567 \mu ^4 \omega ^{10}
        +85633164 \mu ^2 \omega ^{12}
\nonumber \\ &
        -12700800 \omega ^{14}\Big)
    +3449600 \Delta ^8 \omega ^2 \Big(
        -135616 \mu ^{14}
        +4759504 \mu ^{12} \omega ^2
        -63690100 \mu ^{10} \omega ^4
        +411704335 \mu ^8 \omega ^6
        -1345235830 \mu ^6 \omega ^8
\nonumber \\ &
        +2119957831 \mu ^4 \omega ^{10}
        -1372908924 \mu ^2 \omega ^{12}
        +203212800 \omega ^{14}\Big)
    +4480 \Delta ^6 \omega ^4 \Big(
        102146880 \mu ^{14}
        -3592164848 \mu ^{12} \omega ^2
        +48223006204 \mu ^{10} \omega ^4
\nonumber \\ &
        -313372930341 \mu ^8 \omega ^6
        +1033472807954 \mu ^6 \omega ^8
        -1657461145741 \mu ^4 \omega ^{10}
        +1115180164692 \mu ^2 \omega ^{12}
        -164195942400 \omega ^{14}\Big)
\nonumber \\ &
    +96 \Delta ^4 \omega ^6 \Big(
        -283275456 \mu ^{14}
        +10153353488 \mu ^{12} \omega ^2
        -140346196932 \mu ^{10} \omega ^4
        +955273089963 \mu ^8 \omega ^6
        -3398941352462 \mu ^6 \omega ^8
\nonumber \\ &
        +6190128382659 \mu ^4 \omega ^{10}
        -5344783028460 \mu ^2 \omega ^{12}
        +780337152000 \omega ^{14}\Big)
    +4 \Delta ^2 \mu ^2 \omega ^8 \Big(
        1661793984 \mu ^{12}
        -57125755280 \mu ^{10} \omega ^2
\nonumber \\ &
        +739137628644 \mu ^8 \omega ^4
        -4506238654467 \mu ^6 \omega ^6
        +13149352856702 \mu ^4 \omega ^8
        -16344830955483 \mu ^2 \omega ^{10}
        +1129207092300 \omega ^{12}\Big)
\nonumber \\ &
    +\mu ^2 \omega ^{10} \Big(
        4565365184 \mu ^{12}
        -157885286352 \mu ^{10} \omega ^2
        +2063250771828 \mu ^8 \omega ^4
        -12803179724687 \mu ^6 \omega ^6
        +38526000508278 \mu ^4 \omega ^8
\nonumber \\ &
        -57376387492551 \mu ^2 \omega ^{10}
        +9431026271100 \omega ^{12}
    \Big)
\Big] ,
\end{align}
\end{subequations}
}
while the respective $Q_{nj}$ polynomials read
{ \footnotesize
\begin{subequations}
\label{eq:Q7J}
\begin{align}
Q_{15} &= 
512 \omega^5 \mu ^3
    \Big( 4\mu^2 -4\omega^2 \Big) 
    \Big( 4\mu^2 - \omega^2 \Big)^2 ,
\\
Q_{17} &= 
16384 \omega^7 \mu ^5
    \Big( 4\mu^2 -9\omega^2 \Big)
    \Big( 4\mu^2 -4\omega^2 \Big)^2 
    \Big( 4\mu^2 - \omega^2 \Big)^3 ,
\\
Q_{19} &= 
6553600 \omega^9 \mu ^7 
    \Big( 4\mu^2 -16\omega^2 \Big) 
    \Big( 4\mu^2 - 9\omega^2 \Big)^2 
    \Big( 4\mu^2 - 4\omega^2 \Big)^3 
    \Big( 4\mu^2 -  \omega^2 \Big)^4 ,
\\
Q_{35} &= 
1024 \omega^5 \mu
    \Big( 4\mu^2 -9\omega^2 \Big)
    \Big( 4\mu^2 -4\omega^2 \Big)
    \Big( 4\mu^2 - \omega^2 \Big) ,
\\
Q_{37} &= 
163840 \omega^7 \mu ^3 
    \Big( 4\mu^2 -16\omega^2 \Big)
    \Big( 4\mu^2 - 9\omega^2 \Big)^2
    \Big( 4\mu^2 - 4\omega^2 \Big)^2
    \Big( 4\mu^2 -  \omega^2 \Big)^2 ,
\\
Q_{39} &= 
1048576000 \omega^9 \mu ^5 
    \Big( 4\mu^2 -25\omega^2 \Big)
    \Big( 4\mu^2 -16\omega^2 \Big)^2
    \Big( 4\mu^2 - 9\omega^2 \Big)^3
    \Big( 4\mu^2 - 4\omega^2 \Big)^3
    \Big( 4\mu^2 -  \omega^2 \Big)^3 ,
\\
Q_{57} &= 
819200 \omega^7 \mu 
    \Big( 4\mu^2 -25\omega^2 \Big)
    \Big( 4\mu^2 -16\omega^2 \Big)
    \Big( 4\mu^2 - 9\omega^2 \Big)
    \Big( 4\mu^2 - 4\omega^2 \Big)
    \Big( 4\mu^2 -  \omega^2 \Big) ,
\\
Q_{59} &= 
18350080000 \omega^9 \mu ^3 
    \Big( 4\mu^2 -36\omega^2 \Big)
    \Big( 4\mu^2 -25\omega^2 \Big)^2
    \Big( 4\mu^2 -16\omega^2 \Big)^2
    \Big( 4\mu^2 - 9\omega^2 \Big)^2
    \Big( 4\mu^2 - 4\omega^2 \Big)^2
%
%
    \Big( 4\mu^2 -  \omega^2 \Big)^2 ,
\\
Q_{79} &= 
10276044800 \omega^9 \mu
    \Big( 4\mu^2 -49\omega^2 \Big)
    \Big( 4\mu^2 -36\omega^2 \Big)
    \Big( 4\mu^2 -25\omega^2 \Big)
    \Big( 4\mu^2 -16\omega^2 \Big)
    \Big( 4\mu^2 - 9\omega^2 \Big)
    \Big( 4\mu^2 - 4\omega^2 \Big)
%
%
    \Big( 4\mu^2 -  \omega^2 \Big) .
\end{align}
\end{subequations}
}
\end{widetext}
%
%
\begin{table*}[t]
\caption[Coefficients for nonlinear response functions]{\label{tab:C1}%
Coefficients $c_{km}^{(nj)}$ for nonlinear response functions.}
\begin{ruledtabular}
\begin{tabular}{lcccccc}
$(n,j,m)$ & $k=0$  & $1$   & $2$   & $3$   & $4$   & $5$\\ \hline
$(1,5,1)$ 
& $10701/32768$     & $34947/8192$
& $-483381/2048$    & $969525/512$
\\
$(1,5,2)$ 
& $-45/32$  & $-153/32$ & $1161/32$ & $-1395/32$
\\
$(1,5,3)$ 
& $38637/32768$ & $14499/8192$
& $-5589/2048$  & $405/512$
\\ \hline
$(1,7,1)$
& $\frac{3600747}{20971520}$ & $-\frac{6782823}{1310720}$
& $\frac{169399269}{131072}$ & $-\frac{651055779}{16384}$
& $\frac{3945678975}{16384}$
\\
$(1,7,2)$
& $ 136161/20480$   & $73161/5120$
& $-323757/2048$    & $905553/1024$
& $-4359915/4096$
\\
$(1,7,3)$
& $-\frac{167924421}{20971520}$ & $-\frac{15788439}{1310720}$
& $3105621/131072$              & $-243027/16384$
& $59535/16384$
\\
$(1,7,4)$
& $ 27/20$      & $261/160$
& $-567/256$    & $189/256$
& $-315/4096$
\\ \hline
$(1,9,1)$
& $-\frac{536495434947}{33554432000}$   &$-\frac{88185569733}{1048576000}$
& $-\frac{4189625617617}{1048576000}$   & $\frac{2120218804413}{5242880}$
& $-\frac{192284854599159}{26214400}$   & $\frac{55631582404839}{1638400}$
\\
$(1,9,2)$   
& $-3050307/81920$              & $190719/40960$
& $-36570717/204800$            & $-17693703/5120$
& $\frac{1775845197}{81920}$    & $-\frac{5334749343}{204800}$
\\
$(1,9,3)$
& $\frac{642944283021}{10737418240}$    & $\frac{249472395639}{2684354560}$
& $-\frac{219108549339}{1677721600}$    & $-\frac{986683977}{83886080}$
& $\frac{1749025089}{41943040}$         & $-\frac{395538633}{52428800}$
\\
$(1,9,4)$ 
& $-58293/8000$     & $-67689/8000$
& $1595349/128000$  & $-14769/2560$
& $268947/204800$   & $-101493/819200$
\\
$(1,9,5)$
& $\frac{2015654625}{2147483648}$   & $\frac{507577995}{536870912}$
& $-\frac{377403543}{335544320}$    & $\frac{6149115}{16777216}$
& $-415611/8388608$                 & $5103/2097152$
\\ \hline
$(3,5,1)$ 
& $9645/65536$  & $14811/16384$
& $-76581/4096$ & $49965/1024$
\\
$(3,5,2)$ 
& $-135/128$    & $-243/128$
& $1107/128$    & $-1305/128$
\\
$(3,5,3)$ 
& $83349/65536$ & $25299/16384$
& $-9549/4096$  & $885/1024$
\\
$(3,5,4)$ 
& $-3/8$    & $-3/8$
& $45/128$  & $-15/256$
\\ \hline
$(3,7,1)$
& $-\frac{53626869}{209715200}$ & $-87837/51200$
& $\frac{823121811}{6553600}$   & $-32860899/16384$
& $\frac{831053307}{163840}$
\\
$(3,7,2)$
& $603219/102400$   & $234537/25600$
& $-1875591/51200$  & $165441/1024$
& $-4318377/20480$
\\
$(3,7,3)$
& $-\frac{12590778429}{1677721600}$ & $-\frac{994320873}{104857600}$
& $\frac{599493141}{52428800}$      & $-\frac{582069}{262144}$
& $611247/1310720$
\\
$(3,7,4)$
& $57/25$           & $3609/1600$
& $-34947/12800$   & $483/512$
& $-4767/40960$
\\
$(3,7,5)$
& $-\frac{29980125}{67108864}$  & $-\frac{1603665}{4194304}$
& $\frac{3655449}{10485760}$    & $-19845/262144$
& $1323/262144$
\\ \hline
$(3,9,1)$
& $-\frac{986591420139}{335544320000}$  & $-\frac{437837072259}{41943040000}$
& $-\frac{4257643835421}{10485760000}$  & $\frac{877567678881}{32768000}$
& $-\frac{76700693056071}{262144000}$   & $\frac{108776910373047}{163840000}$
\\
$(3,9,2)$   
& $-\frac{636647541}{16384000}$ & $-\frac{801618267}{16384000}$
& $\frac{284929479}{8192000}$   & $-\frac{161264883}{327680}$
& $\frac{11771307723}{3276800}$ & $-\frac{628424811}{131072}$
\\
$(3,9,3)$
& $\frac{165604667600259}{3758096384000}$
& $\frac{57586714680177}{939524096000}$
& $-\frac{1054096281783}{16777216000}$  & $-\frac{10238869521}{167772160}$
& $\frac{32777476869}{419430400}$       & $-\frac{82088937}{4194304}$
\\
$(3,9,4)$ 
& $-3703851/896000$             & $-1131831/358400$
& $\frac{6934797}{2560000}$     & $-\frac{3313287}{4096000}$
& $\frac{9481401}{32768000}$    & $-\frac{4836321}{81920000}$
\\
$(3,9,5)$
& $\frac{8794901025}{4294967296}$   & $\frac{1856912067}{1073741824}$
& $-\frac{6048576099}{3355443200}$  & $\frac{470971179}{838860800}$
& $-\frac{6728967}{83886080}$       & $\frac{2366091}{524288000}$
\\
$(3,9,6)$
& $-\frac{24938361}{81920000}$  & $-\frac{18784143}{81920000}$
& $\frac{8067843}{40960000}$    & $-\frac{371763}{8192000}$
& $70119/16384000$              & $-11907/81920000$
\\ \hline
$(5,7,1)$
& $-\frac{42084549}{1048576000}$    & $-\frac{18675387}{65536000}$
& $\frac{244263681}{32768000}$      & $-4560003/163840$
& $10682847/819200$
\\
$(5,7,2)$
& $55539/81920$     & $24921/20480$
& $-46251/8192$     & $37125/4096$
& $-53865/16384$
\\
$(5,7,3)$
& $-\frac{31324401}{20971520}$  & $-152361/81920$
& $421767/131072$               & $-15525/8192$
& $6615/16384$
\\
$(5,7,4)$
& $5463/4000$       & $10593/8000$
& $-49437/32000$    & $2637/5120$
& $-12789/204800$
\\
$(5,7,5)$
& $-\frac{5255505}{8388608}$    & $-\frac{1311021}{2621440}$
& $\frac{14508693}{32768000}$   & $-16353/163840$
& $6111/819200$
\\
$(5,7,6)$
& $238383/2048000$  & $40581/512000$
& $-57267/1024000$  & $189/20480$
& $-189/409600$
\\ \hline
$(5,9,1)$
& $\frac{577353118113}{4110417920000}$  & $\frac{2638802199357}{4110417920000}$
& $-\frac{48508858079019}{1284505600000}$   
& $\frac{14516873633097}{18350080000}$
& $-\frac{ 235727258373}{81920000}$     & $\frac{15796816546743}{11468800000}$
\\
$(5,9,2)$   
& $-\frac {3998224989}{802816000}$  & $-\frac{6288282153}{802816000}$
& $\frac{6780010203}{401408000}$    & $-\frac{935238663}{11468800}$
& $\frac{109778031}{655360}$        & $-\frac{7597555947}{114688000}$
\\
$(5,9,3)$
& $\frac{1729523840823}{268435456000}$  & $\frac{ 3980375587593}{469762048000}$
& $-\frac{462887122311}{58720256000}$   & $-\frac{2456297433}{419430400}$
& $\frac{1020541167}{209715200}$        & $-\frac{178915797}{262144000}$
\\
$(5,9,4)$ 
& $-\frac{9642843}{3200000}$    & $-\frac{108121941}{44800000}$
& $\frac{192500793}{89600000}$  & $-\frac{10960173}{20480000}$
& $\frac{20177523}{163840000}$  & $-\frac{10064979}{409600000}$
\\
$(5,9,5)$
& $\frac{1039550600547}{526133493760}$  & $\frac{5087867850423}{3288334336000}$
& $-\frac{15344124829389}{10276044800000}$
& $\frac{32162212953}{73400320000}$
& $-\frac{329176917}{5242880000}$       & $\frac{180046557}{ 45875200000}$
\\
$(5,9,6)$
& $-\frac{13064400981}{20070400000}$    & $-\frac{9073263033}{20070400000}$
& $\frac{3707954739}{10035200000}$      & $-\frac{4888539}{57344000}$
& $\frac{700947}{81920000}$             & $-\frac{187839}{573440000}$
\\
$(5,9,7)$
& $\frac{6079966893}{67108864000}$  & $\frac{936300519}{16777216000}$
& $-\frac{563601753}{14680064000}$  & $\frac{732699}{104857600}$
& $-27027/52428800$                 & $891/65536000$
\\ \hline 
$(7,9,1)$   
& $\frac{6230389014279}{920733614080000}$   
& $\frac{12156373280043}{230183403520000}$
& $-\frac{43104455194473}{28772925440000}$  
& $\frac{1361662428573}{205520896000}$
& $-\frac{15649268877}{2936012800}$
& $\frac{108596636577}{128450560000}$
\\
$(7,9,2)$   
& $-\frac{1230921}{6553600}$    & $-\frac{15871743}{45875200}$
& $\frac{35218017}{22937600}$   & $-1843497/655360$
& $2153151/1310720$             & $-\frac{1517373}{6553600}$
\\
$(7,9,3)$
& $\frac{219591204687}{375809638400}$   & $\frac{14174937711}{18790481920}$
& $-\frac{16040957553}{11744051200}$    & $\frac{77976621}{83886080}$
& $-\frac{12826863}{41943040}$          & $\frac{1740609}{52428800}$
\\
$(7,9,4)$
& $-991017/1120000$             & $-\frac{1978677}{2240000}$
& $\frac{20060973}{17920000}$   & $-934083/2048000$
& $18459/204800$                & $-\frac{1123551}{163840000}$
\\
$(7,9,5)$
& $\frac{1746312975}{2147483648}$   & $\frac{2487748509}{3758096384}$
& $-\frac{7662410271}{11744051200}$ & $\frac{79887807}{419430400}$
& $-\frac{1070433}{41943040}$       & $\frac{353403}{262144000}$
\\
$(7,9,6)$
& $-\frac{25577684523}{56197120000}$    & $-\frac{17569607859}{56197120000}$
& $\frac{7051200741}{28098560000}$      & $-\frac{45480069}{802816000}$
& $\frac{255609}{45875200}$             & $-\frac{1688121}{8028160000}$
\\
$(7,9,7)$
& $\frac{3075199281}{21474836480}$      & $\frac{15978363687}{187904819200}$
& $-\frac{65786982381}{1150917017600}$  & $\frac{85423473}{8220835840}$
& $-\frac{462501}{587202560}$           & $\frac{113157}{5138022400}$
\\
$(7,9,8)$
& $-2673/137200$    & $-5589/548800$
& $41067/7024640$   & $-7047/8028160$
& $243/4587520$     & $-729/642252800$
\end{tabular}
\end{ruledtabular}
\end{table*}
%
%

\bibliography{iterative}